\documentclass[aps,preprint,onecolumn,showpacs,showkeys,a4paper,10pt]{revtex4}%
\usepackage{amsfonts}
\usepackage{amsmath}
\usepackage{amssymb}
\usepackage[final,dvips]{graphicx}
\usepackage[dvips]{rotating}%
\begin{document}
\title{The cranking formula and the spurious behaviour of the mass parameters}
\author{B. Mohammed-Azizi}
\affiliation{University of Bechar, Bechar, Algeria}
\email{aziziyoucef@voila.fr}

\begin{abstract}
We discuss some aspects of the approach of the mass parameters by means of the
simple cranking model. In particular, it is well known that the numerical
application of this formula is often subject to ambiguities or contradictions.
It is found that these problems are induced by the presence of two derivatives
in the formula. To overcome these problems, we state a useful ansatz and we
develop a number of simple arguments which tend to justify the removal of
these terms. As soon as this is done, the formula becomes simpler and easier
to interpret. In this respect, it is shown how the shell effects affect the
mass parameters. A number of numerical tests help us in our conclusions.

\end{abstract}
\volumeyear{year}
\volumenumber{number}
\issuenumber{number}
\eid{identifier}
\date[Date text]{date}
\received[Received text]{date}

\revised[Revised text]{date}

\accepted[Accepted text]{date}

\published[Published text]{date}

\startpage{1}
\endpage{ }
\maketitle
\volumeyear{ }

\section{Introduction}

Under the assumption of the adiabaticity of the nuclear motion (the shape
variations are slower than the single-particle motion), a collective
hamiltonian can be defined as the sum of the kinetic and potential energy of
deformation \cite{1}:%
\begin{equation}
H_{\text{collective}}=\frac{1}{2}\sum_{i}\sum_{j}D_{ij}\left\{  \beta
_{1},.,\beta_{n}\right\}  \frac{d\beta_{i}}{dt}\frac{d\beta_{j}}{dt}+U\left\{
\beta_{1},.,\beta_{n}\right\}  \label{Hcollective}%
\end{equation}
In this paper we consider only the deformation of the nucleus neglecting thus
the rotational degrees of freedom. The set $\left\{  \beta_{1},.,\beta
_{n}\right\}  $ specifies a set of deformation parameters of the nuclear
surface. They constitute the dynamical variables of the motion. The functions
$D_{ij}\left\{  \beta_{1},.,\beta_{n}\right\}  $ represent the so-called mass
parameters or the tensor of inertia, and $U\left\{  \beta_{1},.,\beta
_{n}\right\}  $ is the potential energy of deformation. Both of these
functions depend on the deformation of the nucleus. These two quantities are
especially important in the study of the dynamic of the nucleus such as the
nuclear fission (lifetime estimates) \cite{1} and the collective levels of the
nucleus \cite{1a}.\newline Usually, the quantity $U\left\{  \beta_{1}%
,.,\beta_{n}\right\}  $ can be evaluated in the framework of the constrained
Hartree-Fock theory or by the phenomenological shell correction method. The
mass parameters $D_{ij}\left\{  \beta_{1},.,\beta_{n}\right\}  $ are often
approximated by the cranking formula \cite{2} or in the self consistent
approaches by other models \cite{3}-\cite{5}.

The present work focuses mainly on the drawbacks of the numerical aspect of
the calculation of the mass parameters on the basis of the cranking formula.
In particular, it is well known that cranking procedure is somewhat delicate
and in some cases can even lead to singular unphysical values for the mass
parameters. One of the goals of the present work is precisely to discuss this
point and to propose a correction to this formula. First and foremost, to our
point of view, it seems that the main mistake is the presence in the formula
of the derivatives of the gap parameter and the Fermi level. Therefore, the
main part of this work will be devoted to the looking for the arguments (or
proofs) which could justify the removing of these terms. On the basis of
theoretical as well as numerical approaches, this survey will also be very
useful to demonstrate the relative importance of the different terms entering
into the calculations and also to show how the shell effects affect precisely
the mass parameters. We realize that we criticize an old and well established
formula, but it is only "the bad" part of the formula which is called into
question, not the formula itself. This "bad part" is the main source of the
problems of the formula. We are intimately convinced that this criticism will
be very useful and will bring new insight to this subject.

\section{The Inglis-Belyaev or cranking formula}

\subsection{The Cranking or Inglis formula for the mass parameters}

The mass (or vibrational) parameters are given by the Inglis formula \cite{1},
\cite{2}:%

\begin{equation}
D_{ij}\left\{  \beta_{1},.,\beta_{n}\right\}  =2\hbar^{2}\sum_{M\neq0}%
\frac{\left\langle O\right\vert \partial\text{ }/\partial\beta_{i}\left\vert
M\right\rangle \left\langle M\right\vert \partial\text{ }/\partial\beta
_{j}\left\vert O\right\rangle }{E_{M}-E_{O}} \label{massparameters}%
\end{equation}
Where $\left\vert O\right\rangle ,\left\vert M\right\rangle $ are respectively
the ground state and the excited states of the nucleus. The quantities
$E_{M},E_{O}$ are the associated eigenenergies. In the independent-particle
model, whenever the state of the nucleus is assumed to be a Slater determinant
(built on single-particle states of the nucleons), the ground state$\left\vert
O\right\rangle $ will be of course the one where all the particles occupy the
lowest states. The excited states $\left\vert M\right\rangle $ will be
approached by the one particle-hole configurations. In that case, Eq.
(\ref{massparameters}) becomes (see appendix \ref{marker1} ):%

\begin{equation}
D_{ij}\left\{  \beta_{1},.,\beta_{n}\right\}  =2\hbar^{2}\sum_{l>\lambda
,k<\lambda}\frac{\left\langle k\right\vert \frac{\partial}{\partial\beta_{i}%
}\left\vert l\right\rangle \left\langle l\right\vert \frac{\partial}%
{\partial\beta_{i}}\left\vert k\right\rangle }{\epsilon_{l}-\epsilon_{k}}
\label{massparameterssingle11}%
\end{equation}
$\lambda$ denotes the Fermi level. The above expression can be simplified
further if the following simple property is used:\newline$\left(  \epsilon
_{k}-\epsilon_{l}\right)  \left\langle k\right\vert \partial$ $/\partial
\beta_{j}\left\vert l\right\rangle =\left\langle k\right\vert \left[
H_{sp},\partial/\partial\beta_{j}\right]  \left\vert l\right\rangle
=-\left\langle k\right\vert \partial H_{sp}/\partial\beta_{j}\left\vert
l\right\rangle $ for $k\neq l$\newline so that Eq.
(\ref{massparameterssingle11}) becomes%

\begin{equation}
D_{ij}\left\{  \beta_{1},.,\beta_{n}\right\}  =2\hbar^{2}%
{\textstyle\sum\limits_{l>\lambda,k<\lambda}}
\frac{\left\langle k\right\vert \frac{\partial H_{sp}}{\partial\beta_{i}%
}\left\vert l\right\rangle \left\langle l\right\vert \frac{\partial H_{sp}%
}{\partial\beta_{i}}\left\vert k\right\rangle }{\left(  \epsilon_{l}%
-\epsilon_{k}\right)  ^{3}} \label{massparameterssingle2}%
\end{equation}
where $H_{sp}$ is the single-particle hamiltonian. The single particle states
are given by the Schrodinger equation of the independent-particle model.%

\begin{equation}
H_{sp}\left\vert \nu\right\rangle =\epsilon_{\nu}\left\vert \nu\right\rangle
\label{schrodinger}%
\end{equation}

\subsection{The Inglis-Belyaev or cranking formula with pairing correlations
\label{eground}}

It must be noted that in Eq. (\ref{massparameterssingle2}) the denominator
$\epsilon_{l}-\epsilon_{k}$ vanishes in the case where the Fermi level
coincides with two or more degenerate levels. This is the major drawback of
the formula. It is possible to overcome this difficulty by taking into account
the pairing correlations. This can be achieved through the BCS\ approximation
by the following replacements in Eq. (\ref{massparameters}):\newline\qquad i)
the ground state $\left\vert O\right\rangle $ by the BCS state $\left\vert
BCS\right\rangle .$\newline\qquad ii) the excited states $\left\vert
M\right\rangle $ by the two quasiparticle excitations states $\left\vert
\mu,\nu\right\rangle $ (here we consider only the even-even nuclei). This led
Belyaev \cite{6} to two types of non-vanishing matrix elements (see also the
demonstration given in the appendix \ref{marker2}):%
\begin{align}
\left\langle \nu,-\mu\right\vert \partial/\partial\beta_{i}\left\vert
BCS\right\rangle  &  =\left(  u_{\nu}\upsilon_{\mu}+u_{\mu}\upsilon_{\nu
}\right)  \left\langle \nu\right\vert \frac{\partial}{\partial\beta_{i}%
}\left\vert \mu\right\rangle \text{ \ \ \ \ }\nu\neq\mu\label{nondiago1}\\
\left\langle \nu,-\nu\right\vert \partial/\partial\beta_{i}\left\vert
BCS\right\rangle  &  =-\frac{1}{\upsilon_{\nu}}\frac{\partial u_{\nu}%
}{\partial\beta_{i}} \label{diago1}%
\end{align}
The later is due to the changing of the occupations probabilities with the
deformation, i.e. with $u_{\nu}$ and $\upsilon_{\nu}$:\newline\qquad iii) the
energy $E_{O\text{ }}$ by $E_{BCS}$ and $E_{M}$ by the energy of the two
quasiparticles, i.e., by $E_{\nu}+E_{\mu}+E_{BCS}$. The quasiparticle energy
being $E_{\nu}=\sqrt{\left(  \epsilon_{\nu}-\lambda\right)  ^{2}+\Delta^{2}}$
and the BCS state is defined from the "true" vacuum $\left\vert 0\right\rangle
$ by: $\left\vert BCS\right\rangle =\Pi_{k}\left(  u_{k}+\upsilon_{k}a_{k}%
^{+}a_{-k}^{+}\right)  \left\vert 0\right\rangle $.\newline The "Belyaev
formulation" for the mass parameters (formula (\ref{a5}) of the appendix
\ref{appendix b}) is then:%

\begin{equation}
D_{ij}\left\{  \beta_{1},.,\beta_{n}\right\}  =2\hbar^{2}%
{\displaystyle\sum_{\nu}}
{\displaystyle\sum_{\mu\neq\nu}}
\dfrac{\left(  u_{\nu}\upsilon_{\mu}+u_{\mu}\upsilon_{\nu}\right)  ^{2}%
}{E_{\nu}+E_{\mu}}\left\langle \nu\right\vert \frac{\partial}{\partial
\beta_{i}}\left\vert \mu\right\rangle \left\langle \mu\right\vert
\frac{\partial}{\partial\beta_{j}}\left\vert \nu\right\rangle +\hbar
^{2}\underset{\nu}{%
{\displaystyle\sum}
}\frac{1}{E_{\nu}}\frac{1}{\upsilon_{\nu}^{2}}\frac{\partial u_{\nu}}%
{\partial\beta_{i}}\frac{\partial u_{\nu}}{\partial\beta_{j}}
\label{massparametersbcs1}%
\end{equation}
Beside this formula, there is an other more convenient formulation due to Bes
\cite{7} modified slightly by the authors of Ref. \cite{1}. In the latter, the
above two types of matrix elements have been cast under a more explicit form
(see again the demonstrations in appendices \ref{appendix c} and
\ref{appendix d}),%
\begin{align}
\left\langle \nu,-\mu\right\vert \partial/\partial\beta_{i}\left\vert
BCS\right\rangle  &  =-\dfrac{u_{\nu}\upsilon_{\mu}+u_{\mu}\upsilon_{\nu}%
}{E_{\nu}+E_{\mu}}\left\langle \nu\right\vert \frac{\partial H_{sp}}%
{\partial\beta_{i}}\left\vert \mu\right\rangle \text{ \ \ \ \ }\nu\neq
\mu\label{nondiago2}\\
\left\langle \nu,-\nu\right\vert \partial/\partial\beta_{i}\left\vert
BCS\right\rangle  &  =-\text{ }\Delta\left\langle \nu\right\vert
\frac{\partial H_{sp}}{\partial\beta_{i}}\left\vert \nu\right\rangle
+\Delta\dfrac{\partial\lambda}{\partial\beta_{i}}+\left(  \epsilon_{\nu
}-\lambda\right)  \dfrac{\partial\Delta}{\partial\beta_{i}} \label{diago2}%
\end{align}
The final formula of the mass parameters takes now the form (cf. formula
(\ref{a8}) of the appendix \ref{appendix d}):%

\begin{equation}
D_{ij}\left\{  \beta_{1},.,\beta_{n}\right\}  =2\hbar^{2}%
{\displaystyle\sum_{\nu}}
{\displaystyle\sum_{\mu\neq\nu}}
\dfrac{\left(  u_{\nu}\upsilon_{\mu}+u_{\mu}\upsilon_{\nu}\right)  ^{2}%
}{\left(  E_{\nu}+E_{\mu}\right)  ^{3}}\left\langle \nu\right\vert
\dfrac{\partial H_{sp}}{\partial\beta_{i}}\left\vert \mu\right\rangle
\left\langle \mu\right\vert \dfrac{\partial H_{sp}}{\partial\beta_{j}%
}\left\vert \nu\right\rangle +2\hbar^{2}\underset{\nu}{%
{\displaystyle\sum}
}\dfrac{\Delta^{2}}{_{8E_{\nu}^{5}}}R_{i}^{\nu}R_{j}^{\nu}
\label{massparametersbcs2}%
\end{equation}
where%
\begin{equation}
R_{i}^{\nu}=-\left\langle \nu\right\vert \frac{\partial H_{sp}}{\partial
\beta_{i}}\left\vert \nu\right\rangle +\dfrac{\partial\lambda}{\partial
\beta_{i}}+\frac{\left(  \epsilon_{\nu}-\lambda\right)  }{\Delta}%
\dfrac{\partial\Delta}{\partial\beta_{i}} \label{rnu}%
\end{equation}
The two quantities of the r.h.s of Eq. (\ref{massparametersbcs2}) are in the
adopted order, the so-called "non-diagonal" and the "diagonal" parts of the
mass parameters. The derivatives are contained in the above diagonal term
$R_{i}^{\nu}$. When the derivatives $\partial\lambda/\partial\beta_{i}$,
$\partial\Delta/\partial\beta_{i}$ cancel the diagonal term reduce simply to
the diagonal matrix element $-\left\langle \nu\right\vert \partial
H_{sp}/\partial\beta_{i}\left\vert \nu\right\rangle $. In order to facilitate
the comparison with other papers, we recall that the cranking formula is
usually cast under a slightly different form:%

\begin{equation}
D_{ij}\left\{  \beta_{1},.,\beta_{n}\right\}  =2\hbar^{2}%
{\displaystyle\sum_{\nu}}
{\displaystyle\sum_{\mu}}
\dfrac{\left(  u_{\nu}\upsilon_{\mu}+u_{\mu}\upsilon_{\nu}\right)  ^{2}%
}{\left(  E_{\nu}+E_{\mu}\right)  ^{3}}\left\langle \nu\right\vert
\dfrac{\partial H_{sp}}{\partial\beta_{i}}\left\vert \mu\right\rangle
\left\langle \mu\right\vert \dfrac{\partial H_{sp}}{\partial\beta_{j}%
}\left\vert \nu\right\rangle +P_{ij} \label{standard}%
\end{equation}
where the quantity $P_{ij}$ enclose all (and only) the derivatives \cite{1}
because the product of matrix element of $R_{i}^{\nu}R_{j}^{\nu}$ is
"displaced" to the first term of the r.h.s. of Eq. (\ref{standard}).

\section{Some precisions on the microscopic model}

\subsection{The Schrodinger equation}

In this work the numerical tests are based on a microscopic hamiltonian
(represented by $H_{sp\text{ }}$in the previous formulae). The latter is
defined starting from a deformed Woods-Saxon potential. In order to obtain a
realistic model, a spin-orbit\ and a Coulomb (protons case) terms are also
taken into account in the hamiltonian.\newline Since in the hamiltonian the
kinetic energy operator does not depend on the deformation and since the
deformed average potential is the most important term compared to the
spin-orbit and coulomb interactions, the derivative of the single-particle
hamiltonian (appearing in the above cranking formula) will be approximated by
the one of the average potential as in Ref.\cite{1}:%
\begin{equation}
\left\langle \nu\right\vert \frac{\partial H_{sp}}{\partial\beta_{i}%
}\left\vert \mu\right\rangle \approx\left\langle \nu\right\vert \frac{\partial
V}{\partial\beta_{i}}\left\vert \mu\right\rangle \label{approx}%
\end{equation}
$V$ denotes the deformed Wood-Saxon potential. In our work, the deformation of
the nuclear surface and hence the one of the mean field $V$ is of the
quadrupole type and is given by the well known Bohr parameters $(\beta
,\gamma)$. These quantities are connected to the elongation and the axial
asymmetry of the nucleus. In this special case, our notation reduces here to
these two deformations parameters:%
\begin{equation}
\left\{  \beta_{1},.,\beta_{n}\right\}  =\left\{  \beta,\gamma\right\}
\label{def}%
\end{equation}
The solution of the eigenvalues problem is obtained as follows:\newline%
\qquad(i) The eigenfunctions $\left\vert \nu\right\rangle $ of this
hamiltonian are expanded onto the basis' functions of the three dimensional
anisotropic harmonic oscillator. The representative matrix of this hamiltonian
is then effectively built in this basis.\newline\qquad(ii) This matrix is then
diagonalized by using a large basis ($\sim16-20$ major shells of the
oscillator basis).\newline One obtains in this way, the single-particle
energies spectrum $\left\{  \epsilon_{\nu}\right\}  $ and the set of the
components of associated eigenfunctions $\left\{  \left\vert \nu\right\rangle
\right\}  $ on the oscillator basis. All the details of our microscopic model
and the corresponding FORTRAN program have been given in Ref. \cite{8}%
.\newline Knowing the single-particle spectrum, the next step is to find the
gap parameter $\Delta$ and the Fermi level $\lambda$ by solving the standard
BCS equations. Technical details of these calculations can be found from Ref
\cite{9}. As soon as $\Delta$ and $\lambda$ are known, it becomes easy to
deduce the BCS amplitudes $\upsilon_{\nu}^{{}}$, $u_{\nu}$ associated with the
energy level $\epsilon_{\nu}$. Thus, at this stage, all the quantities such as
$\epsilon_{\nu},\left\vert \nu\right\rangle ,\Delta,\lambda,u_{\nu}%
,\upsilon_{\mu}$ which are necessary in Eq. (\ref{massparametersbcs2}) are known.

\subsection{Deformation dependence of $\Delta$ and $\lambda$}

Due to the fact that the single-particle hamiltonian $H_{sp\text{ }}$ depends
explicitly on the deformation, its eigenenergies and its eigenfunctions will
be also explicitly deformation dependent. To keep in mind that the deformation
dependence comes only from the Schrodinger equation, it is useful to highlight
this explicit dependence:%
\begin{equation}
H_{sp}(\beta,\gamma)\left\vert \nu;(\beta,\gamma)\right\rangle =\epsilon_{\nu
}(\beta,\gamma)\left\vert \nu;(\beta,\gamma)\right\rangle _{{}}
\label{schrodinger2}%
\end{equation}
Although being explicit, the above notation is somewhat cumbersome, therefore,
for brevity the dependence on the deformation of the above quantities is
usually omitted. \newline Starting from the single-particle spectrum the gap
parameters $\Delta$ and the Fermi level $\lambda$ are solved from the BCS
equations (\ref{bcs1}) and (\ref{bcs2}) as soon as the single-particle
spectrum $\left\{  \epsilon_{\nu}\right\}  $ is known.%
\begin{equation}
\dfrac{2}{G}=\underset{\nu=1}{\overset{N_{P}}{\sum}}\frac{1}{\sqrt{\left(
\epsilon_{\nu}-\lambda\right)  ^{2}+\Delta^{2}}} \label{bcs1}%
\end{equation}%
\begin{equation}
N\text{ or }Z=\underset{\nu=1}{\overset{N_{P}}{\sum}}\left(  1-\frac
{\epsilon_{\nu}-\lambda}{\sqrt{\left(  \epsilon_{\nu}-\lambda\right)
^{2}+\Delta^{2}}}\right)  \label{bcs2}%
\end{equation}
Of course, the deformation dependence of the eigenenergies $\epsilon_{\nu
}(\beta)$ involves the ones of $\Delta$ and $\lambda$. However, this
dependence is not obtained "straightforwardly" from the single particle model
itself, that is, from the Shrodinger equation (\ref{schrodinger2}), but from
the BCS equations (\ref{bcs1}) and (\ref{bcs2}). From this point of view,
these two quantities (i.e. $\Delta$ and $\lambda$) must not be linked to the
deformation as the eigenenergies and eigenfunctions of $H_{sp}$ do.\newline
Nevertheless, usually they are considered "explicitly" deformation dependent
through the expression of their derivatives obtained by the so-called
lowest-order expansion \cite{11}, \cite{1}:%
\begin{gather}
\frac{\partial\lambda}{\partial\beta}=\frac{ac_{\beta}+bd_{\beta}}{a^{2}%
+b^{2}}\text{ \ \ \ \ \ \ \ \ \ }\frac{\partial\lambda}{\partial\gamma}%
=\frac{ac_{\gamma}+bd_{\gamma}}{a^{2}+b^{2}}\label{difflamda}\\
\frac{\partial\Delta}{\partial\beta}=\frac{bc_{\beta}-ad_{\beta}}{a^{2}+b^{2}%
}\text{ \ \ \ \ \ \ \ }\frac{\partial\Delta}{\partial\gamma}=\frac{bc_{\gamma
}-ad_{\gamma}}{a^{2}+b^{2}} \label{diffdelta}%
\end{gather}
with%

\begin{gather}
a=\underset{\nu}{%
{\displaystyle\sum}
}\Delta E_{\nu}^{-3},\text{ \ }b=\underset{\nu}{%
{\displaystyle\sum}
}(\epsilon_{\nu}-\lambda)E_{\nu}^{-3},\label{ab}\\
\text{\ }c_{\beta}=\underset{\nu}{%
{\displaystyle\sum}
}\Delta\left\langle \nu\right\vert \dfrac{\partial H_{sp}}{\partial\beta
}\left\vert \nu\right\rangle E_{\nu}^{-3},\text{ \ }d_{\beta}=\underset{\nu}{%
{\displaystyle\sum}
}(\epsilon_{\nu}-\lambda)\left\langle \nu\right\vert \dfrac{\partial H_{sp}%
}{\partial\beta}\left\vert \nu\right\rangle E_{\nu}^{-3}\label{cbdb}\\
c_{\gamma}=\underset{\nu}{%
{\displaystyle\sum}
}\Delta\left\langle \nu\right\vert \dfrac{\partial H_{sp}}{\partial\gamma
}\left\vert \nu\right\rangle E_{\nu}^{-3},\text{ \ }d_{\gamma}=\underset{\nu}{%
{\displaystyle\sum}
}(\epsilon_{\nu}-\lambda)\left\langle \nu\right\vert \dfrac{\partial H_{sp}%
}{\partial\gamma}\left\vert \nu\right\rangle E_{\nu}^{-3} \label{cgdg}%
\end{gather}
In the following the expression " the derivatives" which will be used many
times in the text means simply the both derivatives given by Eq.
(\ref{difflamda}) and (\ref{diffdelta}).

\section{The problem of the phase transition:\ The Singularity in the mass
parameters or the paradox of the cranking formula\label{phtr}}

The transition between the normal and the superfluid phase affects generally
the spherical magic nuclei under changing deformation \cite{11}. As we shall
see, this phenomenon causes the most serious problem to the mass parameters.
For convenience, in the following, we choose to discuss just only one
parameter, namely $D_{\beta\beta}(\beta,\gamma)$. This does by no means
restrict the conclusions of this study.\newline It is well known that the BCS
equations have non-trivial solutions only above a critical value of the
strength $G$ of the pairing interaction. The trivial solution corresponds
theoretically to the value $\Delta=0$ of an unpaired system. In this case, the
mass parameters given by (\ref{massparametersbcs2}) reduces to the ones of the
formula (\ref{massparameterssingle2}), i.e. the cranking formula of the
independent-particle model. Indeed, when $\Delta=0$ it is clear that the
quasi-particle energies of (\ref{massparametersbcs2}) become $E_{\nu
}=\left\vert \epsilon_{\nu}-\lambda\right\vert $, and the quantities $u_{\nu
},\upsilon_{\nu}$ are then either $0$ or $1$ so that the non-diagonal part of
the right hand side of this formula reduces to:%

\begin{align}
&  \lim_{\Delta\rightarrow0}2\hbar^{2}%
{\displaystyle\sum_{\nu}}
{\displaystyle\sum_{\mu\neq\nu}}
\dfrac{\left(  u_{\nu}\upsilon_{\mu}+u_{\mu}\upsilon_{\nu}\right)  ^{2}%
}{\left(  E_{\nu}+E_{\mu}\right)  ^{3}}\left\langle \nu\right\vert
\frac{\partial H_{sp}}{\partial\beta}\left\vert \mu\right\rangle \left\langle
\mu\right\vert \frac{\partial H_{sp}}{\partial\beta}\left\vert \nu
\right\rangle \label{lim}\\
&  =2\hbar^{2}%
{\displaystyle\sum_{\nu>\lambda}}
{\displaystyle\sum_{\mu<\lambda}}
\dfrac{1}{\left(  \epsilon_{\nu}-\epsilon_{\mu}\right)  ^{3}}\left\langle
\nu\right\vert \frac{\partial H_{sp}}{\partial\beta}\left\vert \mu
\right\rangle \left\langle \mu\right\vert \frac{\partial H_{sp}}{\partial
\beta}\left\vert \nu\right\rangle \nonumber
\end{align}
which is the same formula that the one given by Eq.
(\ref{massparameterssingle2}). This implies the important fact that in this
limit ($\Delta\rightarrow0$), the diagonal part (i.e. the second term) of the
right hand side of Eq. (\ref{massparametersbcs2}) must vanish. However in a
few cases for which $\Delta\approx0$, contrary to all expectations, it happens
that this diagonal term leads to very large numerical values near some
"critical deformation".This well-known singular behaviour constitutes
undoubtedly unphysical and undesirable effects. This is the main paradox of
the formula.\newline Because the diagonal matrix element $\left\langle
\nu\right\vert \partial H_{sp}/\partial\beta_{i}\left\vert \nu\right\rangle $
are finite and relatively small (cf. to the numerical examples given below),
it is immediately clear from Eq. (\ref{rnu}) that it is the derivatives of
$\Delta$ and $\lambda$ which cause the problem. In this respect, the formulae
(\ref{difflamda}) and (\ref{diffdelta}) are subject to a major drawback due to
the fact that their common denominator can vanish. This can be easily
explained because on the one hand, $a$ is proportional to $\Delta$ (see Eq.
(\ref{ab})) and therefore vanishes with it and because on the other hand $b$
is defined as a "random" sum of postive and negative values (see again Eq.
(\ref{ab})) depending on whether the terms are below or above the Fermi level.
In the literature, this problem has been reported many times \cite{1},
\cite{10}-\cite{14}, but no solution has been proposed. The authors of Ref.
\cite{1} and \cite{11} claim that for sufficiently large pairing gaps $\Delta$
the total mass parameter is essentially given by the diagonal part without the
derivatives, whereas those of Ref. \cite{14} affirm that the role of the
derivatives is by no mean small in the fission process. Other studies
\cite{13} neglect the derivatives without any justification. Even if in some
cases the derivatives introduce small differences, in others, this is not true
at all. In fact, it is easy to realize that these terms involve serious and
insoluble problems. Therefore, after intensive numerical calculations, we have
led to ask ourselves if their presence could be called into question. If so,
their removal should be justified. Consequently, the problem amounts to find
good arguments for that. This constitutes the first step of this work. Without
further ado, let us examine this point in the next section.

\section{Ansatz and correction of the cranking formula of the mass parameters}

To overcome "the paradox" of the previous section, we need to start from an
ansatz or assertion which justify the removal of the derivatives of $\lambda$
and $\Delta$. \ Before that, we want to recall some important points. First,
the derivatives of $\lambda$ and $\Delta$ come from the derivatives of the
probabilities in the formula (\ref{massparametersbcs1}). It is then worth to
recall their expressions:%
\begin{align}
u_{k}  &  =\frac{1}{\sqrt{2}}\left(  1+\frac{\epsilon_{k}-\lambda}%
{\sqrt{\left(  \epsilon_{k}-\lambda\right)  ^{2}+\Delta^{2}}}\right)
^{1/2}\label{xx1}\\
\upsilon_{k}  &  =\frac{1}{\sqrt{2}}\left(  1-\frac{\epsilon_{k}-\lambda
}{\sqrt{\left(  \epsilon_{k}-\lambda\right)  ^{2}+\Delta^{2}}}\right)  ^{1/2}
\label{xx2}%
\end{align}
These quantities depend implicitly on the deformation $\beta$ through the
level $\epsilon_{k}(\beta)$ of the spectrum and also through $\lambda$ \ and
$\Delta$ which themselves depend implicitly on the entire spectrum $\left\{
\epsilon_{0}(\beta),\epsilon_{1}(\beta),...\right\}  $ via the non linear
equations of the BCS theory. First, we will proceed from a point of view which
somewhat looks like the Virtual works principle. The following ansatz contains
two points:\newline1) Thus, we suppose a "virtual change in deformation"
$\delta\beta$ in such a way that one and only level undergoes change, namely
the level $\epsilon_{k}$. Therefore the "virtual change" of the probibilities
would be in this case.%
\begin{align}
\delta u_{k}  &  =\frac{\partial u_{k}}{\partial\epsilon_{k}}\frac
{\partial\epsilon_{k}}{\partial\beta}\delta\beta+\frac{\partial u_{k}%
}{\partial\lambda}\frac{\partial\lambda}{\partial\epsilon_{k}}\frac
{\partial\epsilon_{k}}{\partial\beta}\delta\beta+\frac{\partial u_{k}%
}{\partial\Delta}\frac{\partial\Delta}{\partial\epsilon_{k}}\frac
{\partial\epsilon_{k}}{\partial\beta}\delta\beta\label{yy1}\\
\delta\upsilon_{k}  &  =\frac{\partial\upsilon_{k}}{\partial\epsilon_{k}}%
\frac{\partial\epsilon_{k}}{\partial\beta}\delta\beta+\frac{\partial
\upsilon_{k}}{\partial\lambda}\frac{\partial\lambda}{\partial\epsilon_{k}%
}\frac{\partial\epsilon_{k}}{\partial\beta}\delta\beta+\frac{\partial
\upsilon_{k}}{\partial\Delta}\frac{\partial\Delta}{\partial\epsilon_{k}}%
\frac{\partial\epsilon_{k}}{\partial\beta}\delta\beta\label{yy2}%
\end{align}
This is very logical because the changes of $\lambda$ and $\Delta$ are
basically due the change of the deformation, not straightforwardly but through
the change of the single-particle levels. In other words we have to write
$\left(  \partial\lambda/\partial\epsilon_{k}\right)  \left(  \partial
\epsilon_{k}/\partial\beta\right)  $ instead $\left(  \partial\lambda
/\partial\beta\right)  $ because the changes of $\lambda$ and $\Delta$ are
determined from the knowledge of the spectrum \{$\epsilon_{k}$\} which depends
itself on the deformation of the nucleus.Thus from our "principle" the
"virtual change of the probability", i.e. $\delta u_{k}$, is due to only one
level, namely "the corresponding" level $\epsilon_{k}$.\newline2) It is well
known from the BCS\ theory that the quantities $u_{k}$, $\upsilon_{k}$,
$\lambda$, $\Delta$, are related to each other self-consistently so that the
change of $\delta u_{k}$ would depend on its own change trough $\lambda$ and
$\Delta$. For example in the BCS theory we have $\Delta=G\sum u_{k}%
\upsilon_{k}$ and $\delta\Delta$ would depend on the changes $\delta u_{k}$
and $\delta\upsilon_{k}$ which in turn depend on $\delta\Delta$. This
constitutes an illogical consequence. To overcome this problem, we state the
following intuitive assertion which avoid such a problem and hence the
"initial" problem of the derivatives of $\lambda$ and $\Delta$.: "Between two
infinitely close "virtual" deformations $\beta$ and $\beta+\delta\beta$, the
quantities $u_{k}$, $\upsilon_{k}$, $\lambda$ and $\Delta$ must verify the
simple (but not so obvious) following conditions":%
\begin{align}
\frac{\partial u_{k}}{\partial\lambda}\frac{\partial\lambda}{\partial
\epsilon_{k}}+\frac{\partial u_{k}}{\partial\Delta}\frac{\partial\Delta
}{\partial\epsilon_{k}}  &  =0\label{yyyy1}\\
\frac{\partial\upsilon_{k}}{\partial\lambda}\frac{\partial\lambda}%
{\partial\epsilon_{k}}+\frac{\partial\upsilon_{k}}{\partial\Delta}%
\frac{\partial\Delta}{\partial\epsilon_{k}}  &  =0 \label{yyyy2}%
\end{align}
in such a way that Eq. (\ref{yy1}) and (\ref{yy2}) reduce to:%

\begin{align}
\delta u_{k}  &  =\frac{\partial u_{k}}{\partial\epsilon_{k}}\frac
{\partial\epsilon_{k}}{\partial\beta}\delta\beta\label{z1}\\
\delta\upsilon_{k}  &  =\frac{\partial\upsilon_{k}}{\partial\epsilon_{k}}%
\frac{\partial\epsilon_{k}}{\partial\beta}\delta\beta\label{z2}%
\end{align}
Implying that:
\begin{align}
\frac{\partial u_{k}}{\partial\beta}  &  =\frac{\partial u_{k}}{\partial
\epsilon_{k}}\frac{\partial\epsilon_{k}}{\partial\beta}\label{s1}\\
\frac{\partial\upsilon_{k}}{\partial\beta}  &  =\frac{\partial\upsilon_{k}%
}{\partial\epsilon_{k}}\frac{\partial\epsilon_{k}}{\partial\beta} \label{s2}%
\end{align}
Must be\ employed instead
\begin{align}
\frac{\partial u_{k}}{\partial\beta}  &  =\frac{\partial u_{k}}{\partial
\epsilon_{k}}\frac{\partial\epsilon_{k}}{\partial\beta}+\frac{\partial u_{k}%
}{\partial\lambda}\frac{\partial\lambda}{\partial\beta}+\frac{\partial u_{k}%
}{\partial\Delta}\frac{\partial\Delta}{\partial\beta}\label{r1}\\
\frac{\partial\upsilon_{k}}{\partial\beta}  &  =\frac{\partial\upsilon_{k}%
}{\partial\epsilon_{k}}\frac{\partial\epsilon_{k}}{\partial\beta}%
+\frac{\partial\upsilon_{k}}{\partial\lambda}\frac{\partial\lambda}%
{\partial\beta}+\frac{\partial\upsilon_{k}}{\partial\Delta}\frac
{\partial\Delta}{\partial\beta} \label{r2}%
\end{align}
In fact a real change of the deformation implies "all the levels of the
spectrum". Consequently the "virtual" changes $\delta u_{k}$ and
$\delta\upsilon_{k}$ must be applied "successively" to all the levels
($u_{k}=u_{1},u_{2,}....$). We recall the important fact that Eq. (\ref{r1})
is used in appendix \ref{appendix d} to obtain the formula
(\ref{massparametersbcs2}) whereas Eq. (\ref{s1}) leads to the following more
simple result (which consists of simply ignoring the derivatives of the "old" formula):%

\begin{equation}
D_{\beta\beta}=2\hbar^{2}\sum_{\nu}\sum_{\mu\neq\nu}\tfrac{\left(  u_{\nu
}\upsilon_{\mu}+u_{\mu}\upsilon_{\nu}\right)  ^{2}}{\left(  E_{\nu}+E_{\mu
}\right)  ^{3}}\left\langle \nu\right\vert \frac{\partial H_{sp}}%
{\partial\beta}\left\vert \mu\right\rangle \left\langle \mu\right\vert
\frac{\partial H_{sp}}{\partial\beta}\left\vert \nu\right\rangle +2\hbar
^{2}\sum_{\nu}\sum_{\mu\neq\nu}\dfrac{\Delta^{2}}{_{8E_{\nu}}}\left\langle
\nu\right\vert \frac{\partial H_{sp}}{\partial\beta}\left\vert \nu
\right\rangle ^{2} \label{massparametersbcs3}%
\end{equation}
In this case, it is very important to note that Eq. (\ref{massparametersbcs3})
reduces to Eq.(\ref{massparameterssingle2}) of an unpaired system and removes
the previous "paradox" of the "old" formula.

Furthermore, for convenience it is to be noted that it is possible to collect
the both terms in the right hand side of the above formula, say, the
non-diagonal and the diagonal parts in only one term. Indeed, with the
identity $2u_{\nu}\upsilon_{\nu}=\Delta E_{\nu}$, we note that for $\nu=\mu$:%
\begin{equation}
\frac{\left(  u_{\nu}\upsilon_{\mu}+u_{\mu}\upsilon_{\nu}\right)  ^{2}%
}{\left(  E_{\nu}+E_{\mu}\right)  ^{3}}=\frac{\left(  \Delta E_{\nu}\right)
^{2}}{\left(  2E_{\nu}\right)  ^{3}}=\frac{\Delta^{2}}{_{8E_{\nu}}}
\label{coeff}%
\end{equation}
so that the missing term in the "non-diagonal part" is precisely the second
contribution of the right hand side of Eq. (\ref{massparametersbcs3}).
Consequently Eq.(\ref{massparametersbcs3}) can be cast under the more compact
form:%
\begin{equation}
D_{ij}(\beta)=2\hbar^{2}\sum_{\nu}\sum_{\mu}c_{\nu\mu}\text{ }\left\langle
\nu\right\vert \tfrac{\partial H_{sp}}{\partial\beta_{i}}\left\vert
\mu\right\rangle \left\langle \mu\right\vert \tfrac{\partial H_{sp}}%
{\partial\beta_{j}}\left\vert \nu\right\rangle \label{massparameters4}%
\end{equation}
with%
\begin{equation}
c_{\nu\mu}=\tfrac{\left(  u_{\nu}\upsilon_{\mu}+u_{\mu}\upsilon_{\nu}\right)
^{2}}{\left(  E_{\nu}+E_{\mu}\right)  ^{3}} \label{coeff2}%
\end{equation}
Thus, the only difference between the two versions of the mass parameters Eq.
(\ref{massparametersbcs2}) and Eq. (\ref{massparametersbcs3}) or Eq.
(\ref{massparameters4}) is the absence of the derivatives $\partial
\lambda/\partial\beta_{i}$ and $\partial\Delta/\partial\beta_{i}$ in the
second version of the formula. The latter constitutes for us the cranking
formula without derivatives or the "corrected" formula. In the following
$c_{\nu\mu}$ will be called simply "the coefficient", and for $\nu=\mu$,
$\left\langle \nu\right\vert \partial H_{sp}/\partial\beta_{i}\left\vert
\mu\right\rangle ^{2}$ will be referred to as "the squared matrix element".

\section{Numerical examples and other arguments in favour of the "corrected"
cranking formula (i.e., without derivatives)}

\subsection{Comparisons between the two variants of the formula for
$\Delta\approx0\label{subo}$}

\begin{figure}[ptb]
\includegraphics[angle=0,width=140mm,keepaspectratio]{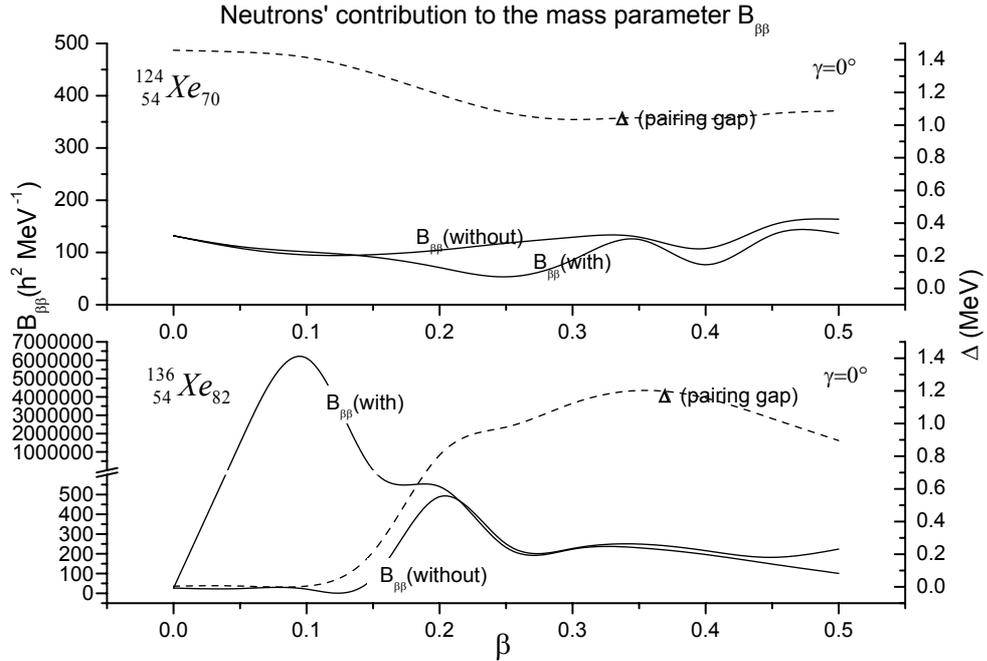}\caption{(Bottom)
Neutron contribution to the mass parameters $B_{\beta\beta}$ for the magic
nucleus $_{54}^{136}Xe_{82}$; The calculations are performed within the
cranking formula including the derivatives and for the same formula without
derivating; Note the quasi divergence of the version with derivatives near the
deformation $\beta=0.1$; The curve without derivatives is characterized by
three region, the first with very small mass parameters (no BCS solution), the
second with strong ( but finite) values with a maximum at $\beta=0.2$, and
from $\beta=0.3$ a mean values about $100-250\hbar^{2}MeV^{-1}$; (Top) same as
(Bottom) but for "mid-shell's neutron" number, i.e., for the nuclei
$_{66}^{136}Xe_{70}$; Note here, the regular behaviour of the pairing gap and
hence the one of the mass parameters $\left(  B_{\beta\beta}\sim
50-150\hbar^{2}MeV^{-1}\right)  $.}%
\label{fig1}%
\end{figure}

In the present section we will try to prove that the modified or "corrected"
cranking formula for the mass parameters is consistent with "reasonable"
results. In this respect, we have first looked for a "critical" case where
there is no BCS solution apart from the trivial solution ($\Delta=0$) and
applied then the both formulae. Such situation is obtained generally for a
magic neutron or proton number. Here, we use the nuclei $_{54}^{136}Xe_{82}$
for which we have plotted in fig. \ref{fig1}-bottom the neutrons' contribution
($N=82=magic$) to the mass parameters vs the quadrupole deformation $\beta$ in
the two versions of the formula.\newline In these calculations the deformation
parameter of the axial asymmetry is fixed to the value $\gamma=0%
{{}^\circ}%
$ (prolate shape). The mention "with" between parentheses signifies that the
calculations are performed with the formula (\ref{massparametersbcs2})
including $\partial\lambda/\partial\beta_{i}$ and $\partial\Delta
/\partial\beta_{i}$, and by "without" the same ones done with the formula Eq.
(\ref{massparametersbcs3}) without these terms. From this example, it is clear
that whenever the gap parameter $\Delta$ becomes small (here $\Delta
\lesssim0.2$ $MEV$) the two formula give very different results for the mass
parameters. The major difference is reached for $\beta\sim0.1$ (for which
$\Delta\approx0.004$) where we obtain values about $6.000000$ $MEV\hbar^{-1}%
$and $25MEV\hbar^{-1}$for the two versions of the formula of the mass
parameter $D_{\beta\beta}$. This singular behaviour (of course within the
version including the derivatives) has obviously nothing to do with reasonable
physical values for the mass parameters. However, both formula give close,
though non-identical, results as soon as $\Delta>0.2$ $MEV$. Thus, this simple
numerical test confirms that it is the derivatives which appear in the formula
(\ref{massparametersbcs2}) that are responsible of the divergence character of
the mass parameters for the case where the valid BCS solution breaks
down.\newline More precisely, we have checked that it is the common
denominator in Eq. (\ref{difflamda}) and (\ref{diffdelta}) which causes the
problem. Indeed, in this denominator the quantities $a$ and $b$ defined by
(\ref{ab}) vanish simultaneously. On this point, we give in the following
table further details concerning the numerical values of $a$, $b$, $c_{\beta}%
$, $d_{\beta}$ which intervene in the quotients of the derivatives in Eq.
(\ref{difflamda}), and Eq. (\ref{diffdelta}).

\begin{table}[ptbh]
\caption{Some numerical details concerning the calculations of the derivatives
given by the formulae (\ref{difflamda}) and (\ref{diffdelta}).}%
\label{t1}%
$%
\begin{tabular}
[c]{|c|c|c|c|c|c|c|c|c|}\hline
$\beta$ & $\gamma$ & $a(MEV^{-2})$ & $b(MEV^{-2})$ & $c_{\beta}(MEV^{-1})$ &
$d_{\beta}(MEV^{-1})$ & $\Delta(MEV)$ & $\frac{\partial\Delta}{\partial\beta
}(MEV)$ & $\frac{\partial\lambda}{\partial\beta}(MEV)$\\\hline\hline
$0.1$ & $0^{{{}^{\circ}}}$ & $0.0096$ & $0.0034$ & $0.0190$ & $-25.7134$ &
$0.0044$ & $2366.3$ & $-832.6$\\\hline
\end{tabular}
$\end{table}Of course, it is to be noted that numerically $\Delta$ does not
vanish rigorously for the trivial solution. This is due to the occupations
probabilities which are numerically not exactly equal to $0$ or $1$. As
mentioned before, the large values of the derivatives are due to the
denominator $a^{2}+b^{2}$ which is very small (For explanations see section
\ref{phtr}). As already noted, this constitutes the main defect of the formula
including the derivatives.On the oher hand, we have also verified that there
is no crossing levels near the Fermi level for this case ($\beta=0.1$). Thus
the derivatives can diverges without any problem of the so-called crossing
levels, as it is often claimed \cite{11}.

\subsection{Single-particle contributions to the "element" $R_{\nu}^{i}$}

For a valid BCS solution (i.e. for $\Delta\gg G$, or approximately
$\Delta\succsim0.80$ in our examples) the problem of the divergence
disappears. Indeed, from fig \ref{fig1}-bottom as soon as $\beta$ exceeds
$0.3$ the gap parameter reaches its normal value and the divergence is no more
present, the both formulae with and without derivatives give close
results.\newline Unlike the closed shell, the mid-shell regions (neutrons or
protons) are characterized by a very regular behaviour of the pairing $\Delta$
and hence of the mass parameters. This is the case in fig \ref{fig1}-top for
the neutron number $N=70$ of the nuclei $_{54}^{136}Xe_{70}$. However, as in
the preceding example, the both formulae (with and without derivatives) do not
give the same results). This is entirely due to the derivatives contained in
the expression of $R_{i}^{\nu}$ which is defined from Eq. (\ref{rnu}). Indeed,
the quantity, $R_{i}^{\nu}$ is the contribution of three terms which are
summarized with their most important corresponding coefficient in the table
(\ref{t2}). \begin{table}[ptbh]
\caption{The most important single-particle contributions to the term
$R_{i}^{\nu}$ given by the formula (\ref{rnu}). Note that the derivative
$\partial\lambda/\partial\beta$ is not so small compared to the diagonal
matrix elements of the derivative (cf. text).}%
\label{t2}%
\[%
\begin{tabular}
[c]{|c|c|c|c|c|}\hline
\multicolumn{5}{|c|}{$\beta=0.50,$ $\gamma=0{{}^{\circ}},$ $\Delta=0.90$ $MeV
$}\\\hline
& coeff. & first contrib. & secd contrib. & third contrib.\\\hline
level number $\nu$ & $2\hbar^{2}c_{\nu\nu}$ & $-$ $\left\langle \nu\right\vert
\frac{\partial H_{sp}}{\partial\beta_{i}}\left\vert \nu\right\rangle $ &
$\dfrac{\partial\lambda}{\partial\beta_{i}}$ & $\frac{\left(  \epsilon_{\nu
}-\lambda\right)  }{\Delta}\dfrac{\partial\Delta}{\partial\beta_{i}}%
$\\\hline\hline
& $\left(  \hbar^{2}MeV^{-3}\right)  $ & $\left(  MeV\right)  $ & $\left(
MeV\right)  $ & $\left(  MeV\right)  $\\\hline
$40$ & $0.117$ & $-0.81$ & $7.66$ & $1.03$\\\hline
$41$ & $0.292$ & $-14.19$ & $7.66$ & $0.38$\\\hline
$42$ & $0.186$ & $-14.50$ & $7.66$ & $-0.75$\\\hline
$43$ & $0.108$ & $0.72$ & $7.66$ & $-1.07$\\\hline
\end{tabular}
\]
\end{table}The energy levels $\epsilon_{\nu}$ are labeled by $\nu$ according
to an increasing order as already pr\'{e}cised and the associated
eigenfunctions are noted by $\left\vert \nu\right\rangle $. In this particular
case, we have took $N=82$ neutrons. There is thus $41$ pairs of particles in
time-conjugate states. For this reason the Fermi level is close to the
$41^{th}$level, i.e., the last filled level. This explains why the most
important levels are labeled with indices $\nu$ which are close to $41$.From
this table it is clear that the third contribution is small near the Fermi
level, $\partial\Delta/\partial\beta_{i}$ being also relatively small for such
a value of the pairing gap ($\Delta=0.90$ $MeV$). On the other hand, the
contribution $\partial\lambda/\partial\beta_{i}$ which does not depend on the
level $\nu$ is by no means so small ($\sim7.66MeV$) compared to the matrix
element. This result contradicts some works in which the derivatives are
neglected, claiming that the derivatives are small in the case of a valid BCS
state. Further checkings have shown that it is $\partial\Delta/\partial
\beta_{i}$ which is responsible of the singularity ($\Delta\approx0$) whereas
it is $\partial\lambda/\partial\beta_{i}$ which modifies the results for valid
BCS solutions. Thus, in all the cases the derivatives modify the results. In
fact, in terms of physical arguments, we can say that the derivative
$\partial\lambda/\partial\beta_{i}$ perturbs or "masks" the shell effects of
the diagonal matrix element (see the next subsection) and this consitutes an
additional physical argument in favour of the removal of the derivatives.

\subsection{Matrix elements: comparison between the diagonal and the
non-diagonal parts. The shell effects lie in the diagonal part}

\begin{figure}[ptb]
\includegraphics[angle=0,width=140mm,keepaspectratio]{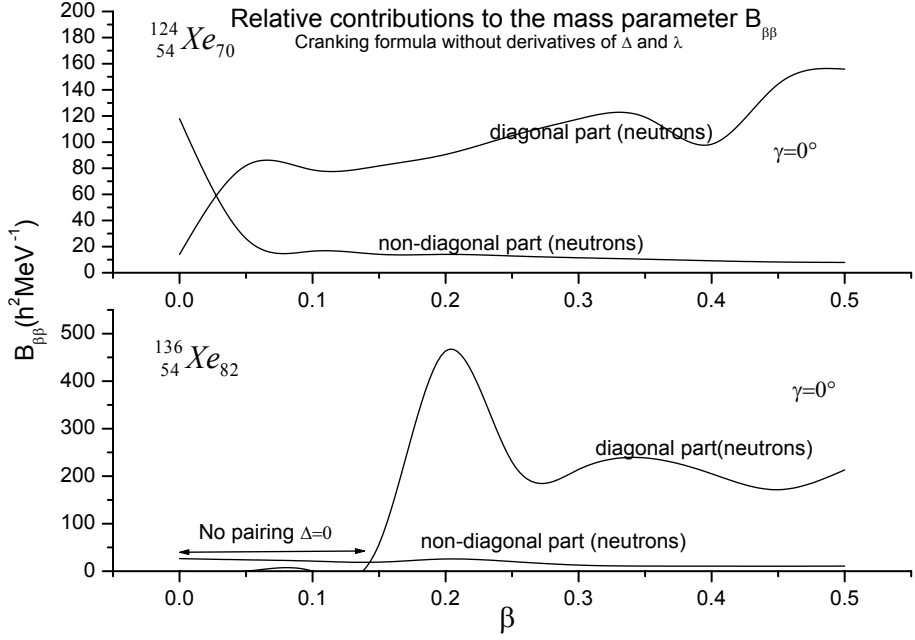}\caption{(Bottom
and Top) same as fig 1; Here, the figure details the diagonal contribution
without derivatives and the non-diagonal one; Apart from the spherical case
($\beta=0$) for $N=70$ which constitutes a "special case", the diagonal part
is by far always more important.}%
\label{fig2}%
\end{figure}

At this stage, the effects of the "perturbations" involved by the derivatives
are known. Therefore, from now, we will concentrate only on the formula of the
cranking without derivatives. It is interesting to compare the magnitude of
the contributions of the diagonal and non-diagonal parts given by Eq.
(\ref{massparametersbcs3}). These two contributions are plotted in fig.
\ref{fig2} for a magic number of neutrons $N=82$ (bottom) and for a mid-shell
neutron number $N=70$ (top).\ In the both cases, the diagonal contribution is
dominant (about 10 times larger) if we except the region of the spherical
shape ($\beta=0$) for $N=70$ which seems to be very particular.
Apart from the noted special spherical case, the explanation of
these results can be easily understood. Indeed, first, the diagonal matrix
elements are always by far more important than the other because in the
integrand the derivative of the potential are multiplied by a squared wave
function,%
\begin{equation}
\left\langle \nu\right\vert \tfrac{\partial H_{sp}}{\partial\beta}\left\vert
\nu\right\rangle \approx\left\langle \nu\right\vert \tfrac{\partial
V}{\partial\beta}\left\vert \nu\right\rangle =\int\left\vert \varphi_{\nu
}(\overrightarrow{r},\beta,\gamma)\right\vert ^{2}\tfrac{\partial
V(\overrightarrow{r},\beta,\gamma)}{\partial\beta}d^{3}r \label{diago}%
\end{equation}
whereas for the non-diagonal matrix elements, i.e., for%
\begin{equation}
\left\langle \nu\right\vert \tfrac{\partial H_{sp}}{\partial\beta}\left\vert
\mu\right\rangle \approx\left\langle \nu\right\vert \tfrac{\partial
V}{\partial\beta}\left\vert \mu\right\rangle =\int\varphi_{\nu}^{\ast
}(\overrightarrow{r},\beta,\gamma)\tfrac{\partial V(\overrightarrow{r}%
,\beta,\gamma)}{\partial\beta}\varphi_{\mu}(\overrightarrow{r},\beta
,\gamma)d^{3}r \label{nondiago}%
\end{equation}
the two radial wave functions are not always in phase, their product
$\varphi_{\nu}^{\ast}(\overrightarrow{r},\beta,\gamma)\varphi_{\mu
}(\overrightarrow{r},\beta,\gamma)$ has not a definite sign, the contributions
to the integral will have tendency to cancel each other, or at least to lead
to a small values.\newline It is clair that the more important coefficients
$c_{\nu\mu}$ (diagonal and non diagonal) are those which are the closest to
the Fermi level. Indeed, these coefficients are which given by (see Eq.
(\ref{coeff2})), are maximums when the quasiparticle energy $E_{\nu}%
=\sqrt{\left(  \epsilon_{\nu}-\lambda\right)  ^{2}+\Delta^{2}}$ reaches its
minimal value, i.e. for $\epsilon_{\nu}\approx\epsilon_{\mu}\approx\lambda$.
The few important non-diagonal coefficients are generally multiplied by weak
matrix elements. Indeed, for closed levels ($\nu\approx\mu,$ $\nu\neq\mu$),
the radial wave functions are often in "opposite" phase (like two
"consecutives" wave functions of the one dimensional infinite square well) so
that the contribution to the integral will be necessarily very small. This
explains why the diagonal part is dominant.\newline Before closing this
paragraph, we would like to add an important remark concerning the shell
effects. The latter are of course due to quantum mechanics. They involve
sudden variations in the mass parameters with the number of particles. They
appear when a few states contribute essentially to the mass parameters. In
fact, for the mass parameters, we have a very limited number of diagonal
matrix elements which contribute essentially to the mass parameters opposed to
an admixture of a large number of small non-diagonal matrix elements.
Consequently, it is clear that the shell effects will be contained in the
diagonal part (of course without the derivatives).

\subsection{Magnitude of the different terms\label{mx}}

\begin{figure}[ptb]
\includegraphics[angle=0,width=140mm,keepaspectratio]{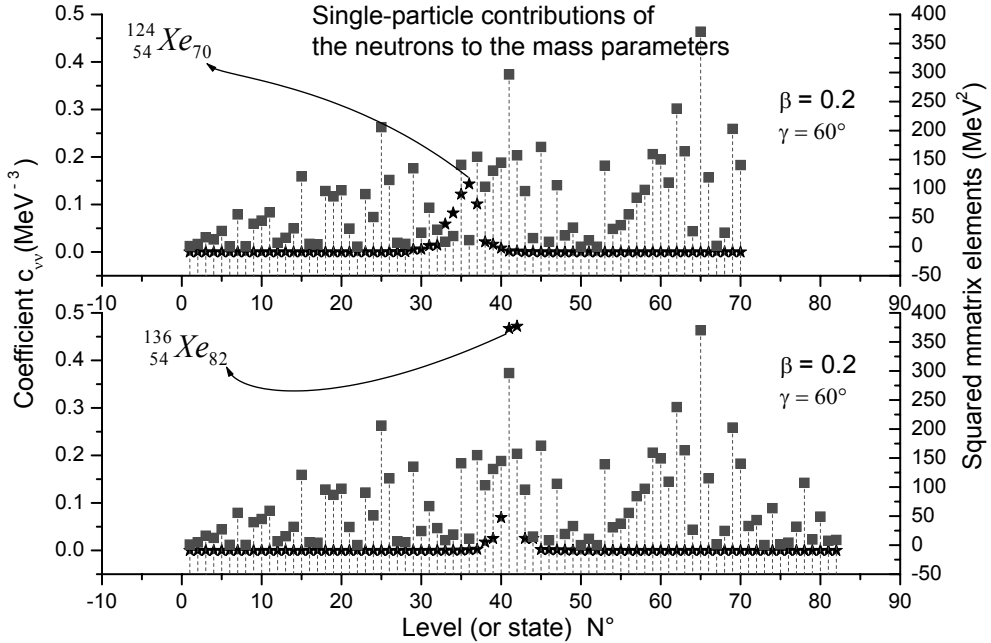}\caption{Following
the same examples as those given in fig (1) and (2), we precise here the
single-particle contributions of the diagonal- part for the both isotopes;
These contributions are made of products of the coefficients ($\bigstar$) by
the square of the diagonal matrix elements ($\blacksquare$) (see Rel.
(\ref{diagonalpart})); The arrow indicate the levels which contribute the most
to the mass parameters; For example in the nucleus $_{54}^{136}Xe_{82}$
(bottom) we have $82$ neutrons. Each level supports two neutrons; The las
filled level is thus the $41^{th}$. Therefore the fermi energy is situated
between the $41^{th}$ and the $42^{th}$ level; In the vicinity of this energy
we see the important values of the coefficients which are represented by
stars; This is essentially due to a crossing level near the fermi energy (see
text); The black squares indicate the values of the corresponding squared
matrix elements; Coefficients as well as matrix elements follows their proper
shell structure; Thus the mass parameters are governed by a "dual" shell
structure (see text); (Top) analog results for the second isotope $_{54}%
^{136}Xe_{70}$ with $N=70$ neutrons; In the latter case,of course, the fermi
level will lie between the $35^{th}$ and the $36^{th}$ level; For this case
the coefficients are less important.}%
\label{fig3}%
\end{figure}

We have found numerically that when there is not BCS solution the diagonal
term (without derivatives) is practically equal to zero ($\sim10^{-6}MEV$) so
that the mass parameters reduce only to the non-diagonal part of the formula.
In this case the mass parameters are very small. Because the values and the
fluctuations of the mass parameters are mostly due to the diagonal
contribution, our numerical study will concern exclusively this term. In the
following, once more, we will concentrate obviously on the version of the
formula without derivatives. The "simple" diagonal part can be cast under the
following form:%
\begin{equation}
D_{\beta\beta}(diagonal\text{ }part)=\sum_{\nu}c_{\nu\nu}\,\left\langle
\nu,\right\vert \frac{\partial V}{\partial\beta_{i}}\left\vert \nu
\right\rangle ^{2} \label{diagonalpart}%
\end{equation}
with,%
\begin{equation}
c_{\nu\nu}=(\hbar^{2}/4)\frac{\text{ }\Delta^{2}}{\left(  \sqrt{\left(
\epsilon_{\nu}-\lambda\right)  ^{2}+\Delta^{2}}\right)  ^{5}}
\label{coeffdiago}%
\end{equation}
As mentioned before, only few (about two up to five at all) energy levels
contribute really to the final value of the mass parameters. Indeed, this
function is peaked at $\lambda$ and due to the power of the denominator of Eq.
(\ref{coeffdiago}) decreases very rapidly so that only few levels give
important coefficients. For a valid BCS solution, a relative weak value of
$\Delta$ gives significant coefficient $c_{\nu\nu}$.For example, we give in
fig. \ref{fig3} (bottom and top) the coefficients $c_{\nu\nu}$ (black stars)
and the square of the matrix elements $\left\langle \nu\right\vert \partial
V/\partial\beta_{i}\left\vert \nu\right\rangle ^{2}$ (black squares) for the
neutron contributions of the same two isotopes as before and the same
deformation. First of course, we especially can check that these coefficients
are important near the Fermi level ($41^{th}$ and $35^{th}$level). We have
thus computed the four major neutron contributions of the energy levels and we
have found $97\%$ and $82\%$ respectively for $N=82$ and $N=70$ cases. These
contributions are not only due to the strongest values of the coefficients
$c_{\nu\nu}$ but also to the "corresponding" matrix elements (see Eq.
(\ref{diagonalpart})).We also can see very clearly the shell structure of the
squared matrix elements which is the same for these two cases (because the
deformation and hence the spectrum are the same in the both cases).
An other remark is that the maxima of the matrix elements are
situated at the levels N$%
{{}^\circ}%
15,25,41,65$ which correspond to the nucleon numbers $30,50,82,130$ which are
magic or nearly magic, but these numbers are obtained for a weak deformed
shape of the nucleus, whereas in the binding energy they match the spherical
shape. Moreover these maxima are immediately preceded or followed by small
values. Thus the shell structure of the matrix elements appears less regular
than the one of the energy levels. As we can see, the shell structure is
"contained" not only in the level density but also in the matrix elements.
Thus the mass parameters are characterized by a complex "dual" shell
structure.\newline At last, as a special case, we must point out the very
strong values of the coefficients (i.e., the stars) in Fig \ref{fig3}-bottom
involving very large the mass parameters at $\beta=0.2$. This very exceptional
situation arises in the so called one crossing levels (see Ref.(\cite{10}) and
(\cite{11})) near the Fermi energy. This will be analyzed in the next subsection.

\subsection{Effects of the crossing levels on the diagonal part}

\begin{figure}[ptb]
\includegraphics[angle=0,width=140mm,keepaspectratio]{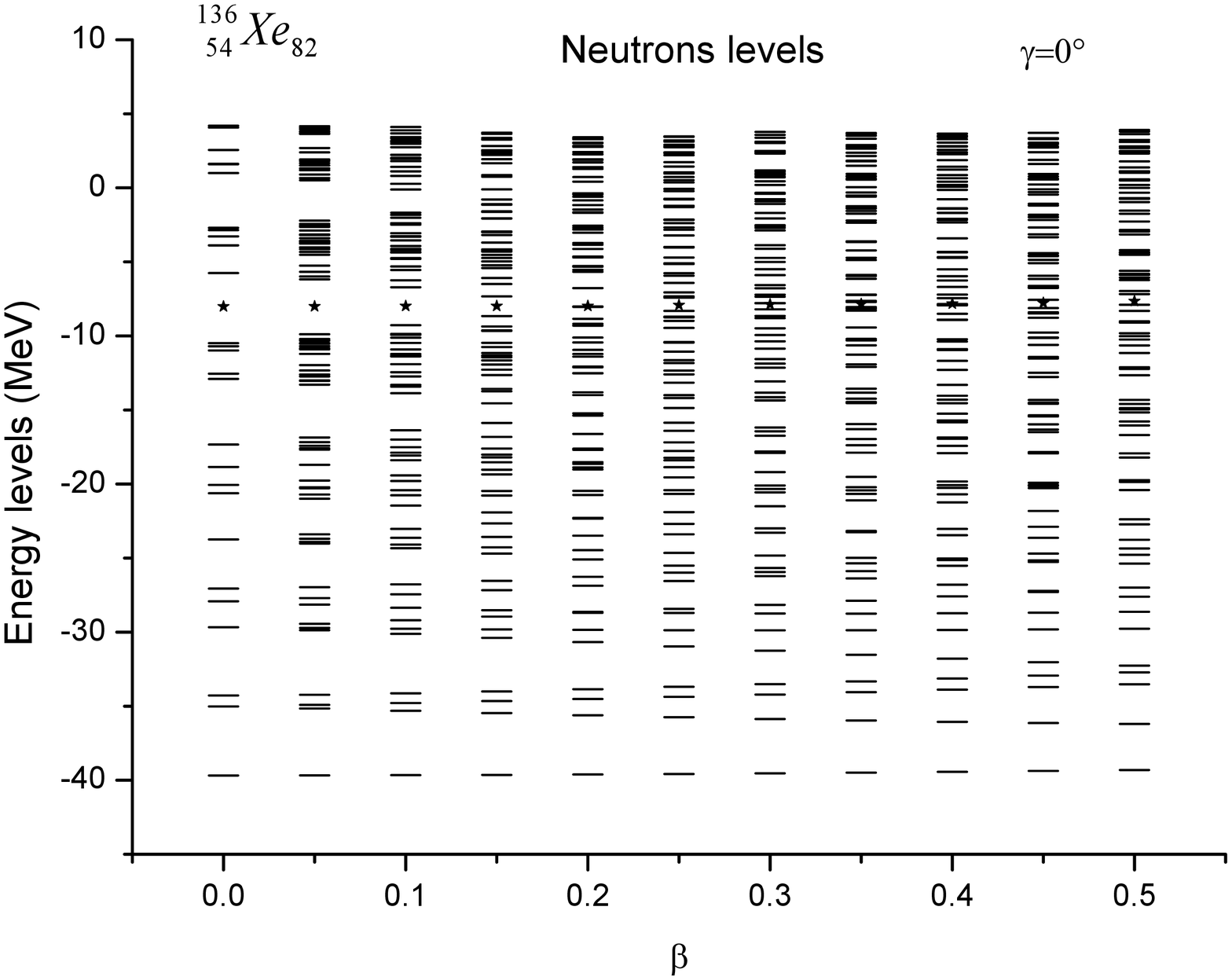}\caption{Energy
levels for $_{54}^{136}Xe_{82}$ at differents deformation; For small
deformations due to the spherical symmetry there is a high degeneracy; For
this reason it seems that ther are less levels at $\beta=0$ and approximately
for the nearest deformations. The fermi level is indicated by a star; Due to
the neutron magic number $N=82,$ it is situated in the center of an important
energy gap (between the $41^{th}$ and the $42^{th}$ level); Note that the gap
reduces gradually as $\beta$ increases from the spherical shape; For the four
first levels there is no BCS solution and the mass parameters are small; For
$\beta=0.2$, we obtain a crossing level which contributes strongly to the mass
parameters through the diagonal part; Compare these comments with the results
of fig(\ref{fig1}-bottom).}%
\label{fig4}%
\end{figure}

In Fig. \ref{fig4}, we have plotted the energy levels for different values of
the deformation parameter $\beta$ ($\gamma$ being fixed to $0$). The BCS Fermi
level is indicated by a star. This level is situated in the middle of a gap
constituted by the last filled and the first unfilled levels. As $\beta$
increases, these two levels approach little by little each other. From the
first $\left(  \beta=0\right)  $ and up to the fourth deformation $\left(
\beta=0.2\right)  $, the Fermi level lies always in the middle of a gap
corresponding to the neutrons magic number $N=82$. In these locations, the level density is so small with respect to
the pairing strength so that it (i.e., the level density) is not able to
insure a valid (non-trivial) BCS solution. In these four cases we have thus
$\Delta\approx0.0044$ $MEV$ and hence no valid BCS solution. Therefore, due to
the weakness of $\Delta$, the diagonal term of (\ref{massparametersbcs3}) does
not contribute at all to the mass parameters. In these situations, the mass
parameters are very small ( $\sim25\hbar^{2}$ $MEV^{-1}$in our case). However,
for the deformation $\beta=0.2$, the two levels become very close to the Fermi
level. It occurs in the immediate vicinity of the Fermi level a crossing
levels. This happens inside a gap and gives rise to two very strong
coefficients. Moreover, the corresponding values of the squared matrix
elements are also important, giving practically all the contribution to the
mass parameters. This phenomenon is characteristic of a phase transition
(compare with \ref{fig1}-bottom). The remaining cases ($\beta$ beyond $0.25$)
can be considered as "normal cases". They correspond to values about
$c_{\nu\nu}\approx0.1-0.2$.

\subsection{Some other important precisions}

We have focused our discussion around the mass parameter $D_{\beta\beta}$, but
similar studies will lead to analog general conclusions (seen above) for the
two other mass parameters $D_{\gamma\gamma}$ and $D_{\beta\gamma}$. However,
we will make further useful remarks. For the quadrupole deformation
$(\beta,\gamma)$, we write the detailed form for the three mass parameters
(naturally, once more without the derivatives):%
\begin{align}
D_{\beta\beta}  &  =2\hbar^{2}\sum\sum c_{\nu\mu}\text{ }\left\langle
\nu\right\vert \tfrac{\partial V}{\partial\beta}\left\vert \mu\right\rangle
^{2}\label{bb}\\
D_{\gamma\gamma}  &  =2\hbar^{2}\sum\sum c_{\nu\mu}\text{ }\left\langle
\nu\right\vert \tfrac{\partial V}{\partial\gamma}\left\vert \mu\right\rangle
^{2}\label{gg}\\
D_{\beta\gamma}  &  =2\hbar^{2}\sum\sum c_{\nu\mu}\text{ }\left\langle
\nu\right\vert \tfrac{\partial V}{\partial\beta}\left\vert \mu\right\rangle
\left\langle \mu\right\vert \tfrac{\partial V}{\partial\gamma}\left\vert
\nu\right\rangle \label{bg}%
\end{align}
with $c_{\nu\mu}$ given by Eq. (\ref{coeff2}).\newline It is easy to compare
these three quantities. First it is to be noted that they have the same
coefficient so that the only difference will come from the matrix elements.
The latter have not a definite sign in such a way that the third mass
parameters will be the result of the sum of positive as well as negative
terms. It will be thus necessary smaller than the two first mass parameters
for which we have only positive contributions. Moreover, it is found
numerically that $D_{\beta\beta}$ is generally larger than $D_{\gamma\gamma}.$
This can also be easily explained by the fact that the single-particle
potential is generally more sensitive to the $\beta$ degree of freedom than
the $\gamma$'s one, especially for the region the Barium isotopes. In other
words for the quadrupole deformations and a Woods-Saxon potential we have,
\begin{equation}
\partial V/\partial\beta\gg\partial V/\partial\gamma\label{dvdb}%
\end{equation}
Although the present work is devoted exclusively to the mass parameters, we
would like to add an important remark concerning the moments of inertia.
Indeed, these quantities can also be evaluated with the help of the cranking
approximation including pairing correlations.(\ref{momentinertia}):%
\begin{equation}
\Im_{k}(\beta,\gamma)=2\hbar^{2}%
{\displaystyle\sum_{\nu}}
{\displaystyle\sum_{\mu}}
\dfrac{\left(  u_{\nu}\upsilon_{\mu}-u_{\mu}\upsilon_{\nu}\right)  ^{2}%
}{E_{\nu}+E_{\mu}}\left\vert \left\langle \mu\right\vert j_{k}\left\vert
\nu\right\rangle \right\vert ^{2} \label{momentinertia}%
\end{equation}
Because the latter looks like to the one of the mass parameter (of course
without the derivatives), it is tempting to conclude that the behaviour of
these two quantities will in principle be the same. However, it is well known
that unlike the mass parameters, the behaviour of the moments of inertia is
always found as a smooth function of particle-number. It depends mainly on the
deformation of the nucleus. This means that contrarily to the mass parameters,
there are no shell effects in the moments of inertia. From our previous study,
it is easy to interpret this difference.

First, this is due partly to the lower power of the denominator of Eq.
(\ref{momentinertia}) but the main reason is that there is no diagonal
contribution ($\nu=\mu$) to this sum because the "coefficient" vanishes when
$\nu=\mu$. Remembering that the shells effects come almost exclusively from
the diagonal part of the total contribution, we can conclude here that they
will be necessary quasi absent in the moments of inertia.

\section{Comparison between the both formulae through the experimental
collective levels}

We have also solved numerically the generalized Bohr Hamiltonian with the help
of the numerical fortran code of \ Ref. \cite{16}. The six inertial functions
have been calculated of course via the two versions of the cranking formula
and the potential energy of deformation with the Strutinsky method. The two
types of calculation are compared to the experimental levels. For this task we
chose two Barium isotopes. This choice is justified by the fact that the magic
nuclei $^{138}Ba_{82}$ undergoes the phase transition near the spherical shape
whereas for the other (non magic) nuclei $^{132}Ba_{76}$ there is not the
case. \begin{figure}[ptb]
\includegraphics[angle=0,width=140mm,keepaspectratio]{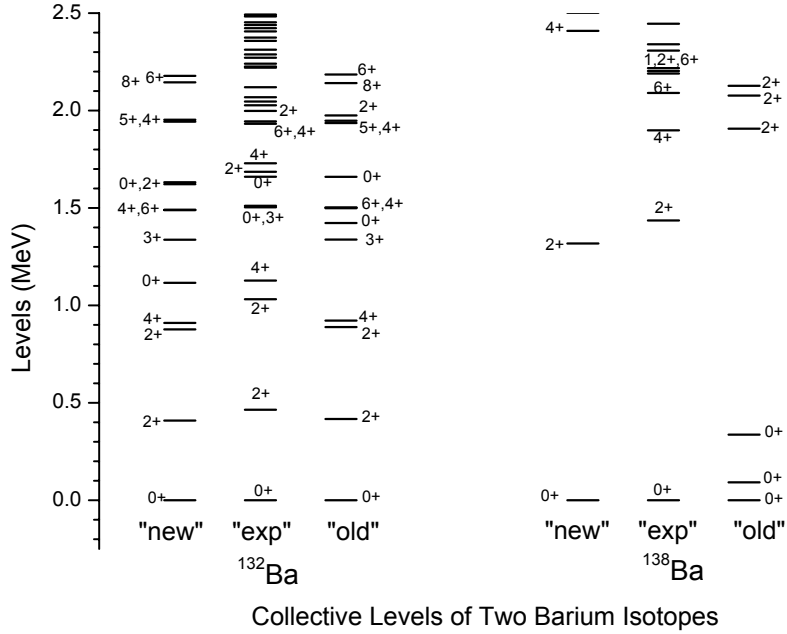}\caption{Collective
levels deduced numerically (see texte) from the generalized Bohr Hamiltonian
including the seven inertial functions (three mass parameters, three moments
of inertia, and the potential energy of the deformation). Two typical cases
with and without phase transition are considered.}%
\label{fig5}%
\end{figure}

The formula without derivatives is called as the "new" formula whereas the
other is called the "old" formula. The experimental low lying collective
levels are denoted by "exp". From the figure it is clear that for the phase
transition $\left(  ^{138}Ba\right)  $ the both formulae give very different
results. The advantage of the new version of the formula is clear because the
old version gives a completely absurd result. In the other case the difference
is much more less important but the theoretical "new" and "old" formulae do
not give identical spectra.

\section{Conclusion}

The cranking formula of the mass parameters is widely used in the study of the
dynamics of the nucleus. However, its applications are not free from
ambiguities and contradictions. The aim of this paper is to convince that
these problems are inherent to a spurious presence of the so-called
derivatives in the formula. Therefore, by means of a simple "proved assertion"
based on a number of pertinent arguments (see text), it is simply recommended
to remove these derivatives. It turns out that the cranking formula without
derivatives is no more subject to the cited problems and becomes simpler to
interpret. Let us re-examine some advantages:

i) The "new" formula reduces in a naturally way to that of the unpaired system
when $\Delta=0$ whereas the old version leads to an inextricable paradox.

ii) The problem of the unphysical large values of the mass (singularity) and
thereby unphysical collective spectra disappears.

iii) The shell effects are physically no more "masked" or "perturbed" by the derivatives.

iv) The formula becomes more transparent and a number a features have been
highlighted in this paper. For example, the shell effects are not only
connected to the level density as it is usually claimed, but also to the
matrix elements which have also their proper shell structure. Thus, the shell
effects appear to be governed by a "dual" shell structure. Therefore,
examining the mass parameters only through the level density, as it is usually
done, is surely insufficient to explain the results.

We hope that this work will help the reader to a better understanding of the
mass parameters\ and will encourage further investigations in this direction.

\appendix

\section{The cranking formula in the independent-particle model\label{marker1}%
\label{appendix a}}

The cranking formula for the mass parameters is given by Eq.
(\ref{massparameters} ). If we want to apply the cranking formula in the
framework of the independent-particle model, we assume that the excited states
are simply one particle-one hole excitations. We proceed to the following replacements:

$\left\vert O\right\rangle \rightarrow\left\vert core\right\rangle $ \ \ and
\ \ $\left\vert M\right\rangle \rightarrow a_{l>\lambda}^{+}a_{k<\lambda}^{{}%
}\left\vert core\right\rangle $ \ \ where \ \ $\left\vert core\right\rangle
=\prod_{n<\lambda}a_{n}^{+}\left\vert 0\right\rangle $ is the ground state or
the core of the independent-particle model. In this expression, $\left\vert
0\right\rangle $ is the "real" vacuum and $a_{n}^{+}$ is the creation operator
of a real particle in the state $n$.\newline In the second quantization
formalism, we can write:\newline$\partial/\partial\beta_{i}=\sum_{\nu,\mu
}\left\langle \nu\right\vert \partial/\partial\beta_{i}\left\vert
\mu\right\rangle a_{\nu}^{+}a_{\mu}^{{}}$\newline We choose $\left\vert
\mu\right\rangle $ as the eigenfunctions of the single particle hamiltonian
$H_{sp}\left\vert \mu\right\rangle =\epsilon_{\mu}\left\vert \mu\right\rangle
$. The $\epsilon_{\mu}$ are then the associated eigenenergies.\newline The
matrix element entering in the cranking formula reads then:\newline%
$\left\langle M\right\vert \partial$ $/\partial\beta_{i}\left\vert
O\right\rangle =\left\langle core\right\vert a_{k<\lambda}^{+}a_{l>\lambda
}^{{}}\sum_{\nu,\mu}\left\langle \nu\right\vert \frac{\partial}{\partial
\beta_{i}}\left\vert \mu\right\rangle a_{\nu}^{+}a_{\mu}^{{}}\left\vert
core\right\rangle $\newline It is clear that the only non-vanishing matrix
elements imply that: $\mu=k$ and $\nu=l$.\newline$\left\langle M\right\vert
\partial$ $/\partial\beta_{i}\left\vert O\right\rangle =\left\langle
l\right\vert \frac{\partial}{\partial\beta_{i}}\left\vert k\right\rangle $ for
$k<\lambda$ and $l>\lambda$. with $\left\langle core\right\vert \left\vert
core\right\rangle =1$.\newline Assuming that in the independent-particle model
the energy of the nucleus amounts simply to the one of its nucleons we will
have $E_{M}-E_{O}=$ $\epsilon_{l}-\epsilon_{k}$, therefore the cranking
formula becomes:%
\begin{equation}
D_{ij}\left\{  \beta_{1},.,\beta_{n}\right\}  =2\hbar^{2}\sum
\limits_{l>\lambda,k<\lambda}\frac{\left\langle k\right\vert \frac{\partial
}{\partial\beta_{i}}\left\vert l\right\rangle \left\langle l\right\vert
\frac{\partial}{\partial\beta_{j}}\left\vert k\right\rangle }{\epsilon
_{l}-\epsilon_{k}} \label{a2}%
\end{equation}

\section{The cranking formula with pairing correlations\label{marker2}%
\label{appendix b}}

The BCS theory gives the ground state and two-quasiparticle excited states for
a paired system as follows \cite{15}:$\left\vert BCS\right\rangle =\prod
_{k>0}(u_{k}+\upsilon_{k}a_{k}^{+}a_{-k}^{+})\left\vert 0\right\rangle
,$\newline$\alpha_{\nu}^{+}\alpha_{\mu}^{+}\left\vert BCS\right\rangle
=a_{\nu}^{+}a_{\mu}^{+}\prod_{k\neq\nu,\mu}(u_{k}+\upsilon_{k}a_{k}^{+}%
a_{-k}^{+})\left\vert 0\right\rangle $ for $\nu\neq\mu,$ and $\alpha_{\nu}%
^{+}\alpha_{-\nu}^{+}\left\vert BCS\right\rangle =(-\upsilon_{\nu}+u_{\nu
}a_{\nu}^{+}a_{-\nu}^{+})\prod_{k\neq\nu}(u_{k}+\upsilon_{k}a_{k}^{+}%
a_{-k}^{+})\left\vert 0\right\rangle $ where $(u_{k}^{{}},\upsilon_{k}^{{}})$
are the usual probability amplitudes of the pairs of particles in the mutual
time-reversed states $\left\vert k\right\rangle $, $\left\vert -k\right\rangle
.$

Starting again from the cranking formula (\ref{massparameters}), we have to
evaluate the matrix elements :$\left\langle M\right\vert \partial$
$/\partial\beta_{i}\left\vert O\right\rangle $. Then, we will replace the
state $\left\vert O\right\rangle $ by the BCS state. The excited states
$\left\vert M\right\rangle $ are supposed to be two quasiparticle excitations
states $\left\vert M\right\rangle =\alpha_{k}^{+}\alpha_{l}^{+}\left\vert
BCS\right\rangle $.\newline Expressing the operator $\partial$ $/\partial
\beta_{i}$ in the second quantization formalism and keeping in mind however
that the differential operator acts not only on the wave functions of the BCS
state but also on the occupations probabilities $u_{k},\upsilon_{k}$, we have
to consider two types of differentiation:

$\frac{\partial}{\partial\beta_{i}}=\left(  \frac{\partial}{\partial\beta_{i}%
}\right)  _{wave\text{ }func}+\left(  \frac{\partial}{\partial\beta_{i}%
}\right)  _{occup.prob}$\newline We must therefore to evaluate successively
two types of matrix elements

\subsection{Calculation of the first type of matrix elements\label{first type}%
}

We have in second quantization representation:\newline$\left(  \frac{\partial
}{\partial\beta_{i}}\right)  _{wave\text{ }func}=\sum_{\nu,\mu}\left\langle
\nu\right\vert \frac{\partial}{\partial\beta_{i}}\left\vert \mu\right\rangle
a_{\nu}^{+}a_{\mu}^{{}}$\newline Applying this operator on the paired system
we find\newline$\left(  \frac{\partial}{\partial\beta_{i}}\right)
_{wave\text{ }func}\left\vert BCS\right\rangle =\sum_{\nu,\mu}\left\langle
\nu\right\vert \frac{\partial}{\partial\beta_{i}}\left\vert \mu\right\rangle
a_{\nu}^{+}a_{\mu}^{{}}\left\vert BCS\right\rangle $\newline Two excited
quasiparticle states of an even-even nucleus are obtained by breaking one pair
of correlated particles so that we must take $\nu\neq\mu$ (It is easy to
verify that this correspond to the first type of the excited states given
above)$.$\newline Using the inverse of the Bogoliubov-Valatin
transformation:\newline$a_{\nu}^{{}}=u_{\nu}^{{}}\alpha_{\nu}-\upsilon_{\nu
}\alpha_{-\nu}^{+}$, \ \ \ \ $a_{-\nu}^{{}}=u_{\nu}^{{}}\alpha_{-\nu}%
+\upsilon_{\nu}\alpha_{\nu}^{+}$\newline in the previous expression, we
find\newline$\sum_{\nu,\mu\neq\nu}\left\langle \nu\right\vert \frac{\partial
}{\partial\beta_{i}}\left\vert \mu\right\rangle (u_{\nu}\upsilon_{\mu}%
\alpha_{\nu}^{+}\alpha_{-\mu}^{+}+.not.two.qp)\left\vert BCS\right\rangle
$\newline We considered here only the states with two quasiparticle (for the
even-even nuclei). We calculate then the first type of matrix
elements:\newline$I_{1}=\left\langle M\right\vert \sum_{\nu,\mu\neq\nu
}\left\langle \nu\right\vert \frac{\partial}{\partial\beta_{i}}\left\vert
\mu\right\rangle u_{\nu}\upsilon_{\mu}\alpha_{\nu}^{+}\alpha_{-\mu}%
^{+}\left\vert BCS\right\rangle $\newline The above form of the formula
suggests that the excited states must be of the form $\left\vert
M\right\rangle =\alpha_{k}^{+}\alpha_{-l}^{+}\left\vert BCS\right\rangle
=\left\vert k,-l\right\rangle $.\newline We obtain then:\newline%
$I_{1}=\left\langle BCS\right\vert \alpha_{-l}^{{}}\alpha_{k}^{{}}\sum
_{\nu,\mu\neq\nu}\left\langle \nu\right\vert \frac{\partial}{\partial\beta
_{i}}\left\vert \mu\right\rangle u_{\nu}\upsilon_{\mu}\alpha_{\nu}^{+}%
\alpha_{-\mu}^{+}\left\vert BCS\right\rangle $\newline$=\sum_{\nu,\mu\neq\nu
}\left\langle \nu\right\vert \frac{\partial}{\partial\beta_{i}}\left\vert
\mu\right\rangle u_{\nu}\upsilon_{\mu}\left\langle BCS\right\vert \alpha
_{-l}^{{}}\alpha_{k}^{{}}\alpha_{\nu}^{+}\alpha_{-\mu}^{+}\left\vert
BCS\right\rangle $\newline We use the following usual Fermions anticommutation
relations:\newline$\left\{  \alpha_{k}^{{}},\alpha_{l}^{{}}\right\}  =\left\{
\alpha_{k}^{+},\alpha_{l}^{+}\right\}  =0,$ \ \ \ $\left\{  \alpha_{k}^{{}%
},\alpha_{l}^{+}\right\}  =\delta_{kl}$\newline Thus the quantity between
brakets gives:\newline$\left\langle BCS\right\vert \alpha_{-l}^{{}}\alpha
_{k}^{{}}\alpha_{\nu}^{+}\alpha_{-\mu}^{+}\left\vert BCS\right\rangle
=\left\langle BCS\right\vert \alpha_{-l}^{{}}\left(  \delta_{\nu k}%
-\alpha_{\nu}^{+}\alpha_{k}^{{}}\right)  \alpha_{-\mu}^{+}\left\vert
BCS\right\rangle $\newline$=\delta_{l\mu}\delta_{\nu k}-\delta_{-\mu k}%
\delta_{-\nu l}$\newline consequently:\newline$\sum_{\nu,\mu\neq\nu
}\left\langle \nu\right\vert \frac{\partial}{\partial\beta_{i}}\left\vert
\mu\right\rangle u_{\nu}\upsilon_{\mu}\left(  \delta_{l\mu}\delta_{\nu
k}-\delta_{-\mu k}\delta_{-\nu l}\right)  $\newline$=\left\langle k\right\vert
\frac{\partial}{\partial\beta_{i}}\left\vert l\right\rangle u_{k}\upsilon
_{l}-\left\langle -k\right\vert \frac{\partial}{\partial\beta_{i}}\left\vert
-l\right\rangle u_{-l}\upsilon_{-k}$\newline the condition $\mu\neq\nu$
implies $k\neq l$ otherwise the result cancels for $k=l.$\newline Noting that
if $T$ is the time-reversal conjugation operator\newline$\left\langle
-k\right\vert \frac{\partial}{\partial\beta_{i}}\left\vert -l\right\rangle
=\left\langle k\right\vert T^{+}\frac{\partial}{\partial\beta_{i}}T\left\vert
l\right\rangle =\left\langle k\right\vert T^{-1}\frac{\partial}{\partial
\beta_{i}}T\left\vert l\right\rangle $\newline and assuming that
$\frac{\partial}{\partial\beta_{i}}$ is time-even, we get:\newline%
$\left\langle -k\right\vert \frac{\partial}{\partial\beta_{i}}\left\vert
-l\right\rangle =\left\langle k\right\vert \frac{\partial}{\partial\beta_{i}%
}\left\vert l\right\rangle $\newline Moreover, using the well-known
property\newline$u_{-l}=u_{l}$, $\upsilon_{-k}=-\upsilon_{k}$\newline we
obtain finally:%

\begin{equation}
I_{1}=\left\langle k,-l\right\vert \left(  \frac{\partial}{\partial\beta_{i}%
}\right)  _{wave\text{ }func}\left\vert BCS\right\rangle =\left(
u_{k}\upsilon_{l}+u_{l}\upsilon_{k}\right)  \left\langle k\right\vert
\frac{\partial}{\partial\beta_{i}}\left\vert l\right\rangle \left(
1-\delta_{k,l}\right)  \label{a3}%
\end{equation}

\subsection{Calculation of the second type of matrix
elements\label{second type}}

Differentiating the BCS state, we obtain:\newline$\left(  \frac{\partial
}{\partial\beta_{i}}\right)  _{occup.prob}\left\vert BCS\right\rangle
=\sum_{\tau}(\frac{\partial u_{\tau}}{\partial\beta_{i}}+\frac{\partial
\upsilon_{\tau}}{\partial\beta_{i}}a_{\tau}^{+}a_{-\tau}^{+})\prod_{k\neq\tau
}(u_{k}+\upsilon_{k}a_{k}^{+}a_{-k}^{+})\left\vert 0\right\rangle $\newline
This corresponds to the second type of the excited states given above because
for the pair in the state $\tau$ we have a linear combination of the vacuum
state with a state of two particles in conjugate states. The excited states
will be necessarily here, of the following form:\newline$\left\vert
M\right\rangle =\alpha_{m}^{+}\alpha_{-m}^{+}\left\vert BCS\right\rangle
=\left\vert m,-m\right\rangle $\newline We have therefore to
calculate:\newline$I_{2}=\left\langle BCS\right\vert \alpha_{-m}^{{}}%
\alpha_{m}^{{}}\sum_{\tau}(\frac{\partial u_{\tau}}{\partial\beta_{i}}%
+\frac{\partial\upsilon_{\tau}}{\partial\beta_{i}}a_{\tau}^{+}a_{-\tau}%
^{+})\prod_{k\neq\tau}(u_{k}+\upsilon_{k}a_{k}^{+}a_{-k}^{+})\left\vert
0\right\rangle .$\newline It is simple to show that:\newline$\prod_{k\neq\tau
}(u_{k}+\upsilon_{k}a_{k}^{+}a_{-k}^{+})\left\vert 0\right\rangle $
$=(u_{\tau}+\upsilon_{\tau}a_{\tau}^{+}a_{-\tau}^{+})^{-1}\left\vert
BCS\right\rangle $\newline then\newline$I_{2}=\left\langle BCS\right\vert
\alpha_{-m}^{{}}\alpha_{m}^{{}}\sum_{\tau}(\frac{\partial u_{\tau}}%
{\partial\beta_{i}}+\frac{\partial\upsilon_{\tau}}{\partial\beta_{i}}a_{\tau
}^{+}a_{-\tau}^{+})(u_{\tau}+\upsilon_{\tau}a_{\tau}^{+}a_{-\tau}^{+}%
)^{-1}\left\vert BCS\right\rangle $\newline The case $m\neq\tau$ leads to the
cancellation of $I_{2}$. The only non vanishing possibility is given by
$\tau=m.$\newline$I_{2}=\left\langle BCS\right\vert \alpha_{-m}^{{}}\alpha
_{m}^{{}}(\frac{\partial u_{m}}{\partial\beta_{i}}+\frac{\partial\upsilon_{m}%
}{\partial\beta_{i}}a_{m}^{+}a_{-m}^{+})(u_{m}+\upsilon_{m}a_{m}^{+}a_{-m}%
^{+})^{-1}\left\vert BCS\right\rangle $\newline Using the inverse
transformation of Bogoliubov-Valatin by "selecting" only two quasiparticle
states\newline$I_{2}=\left\langle BCS\right\vert \alpha_{-m}^{{}}\alpha
_{m}^{{}}(\frac{\partial u_{m}}{\partial\beta_{i}}+\frac{\partial\upsilon_{m}%
}{\partial\beta_{i}}u_{m}^{2}\alpha_{m}^{+}\alpha_{-m}^{+})(u_{m}+\upsilon
_{m}u_{m}^{2}\alpha_{m}^{+}\alpha_{-m}^{+})^{-1}\left\vert BCS\right\rangle
$\newline Making an expansion to first order in $\alpha_{m}^{+}\alpha_{-m}%
^{+}$ of the inverse operator of the above formula (other terms containing
more than two quasiparticle excitations are of course neglected)\newline%
$(u_{m}+\upsilon_{m}u_{m}^{2}\alpha_{m}^{+}\alpha_{-m}^{+})^{-1}=u_{m}%
^{-1}(1+\upsilon_{m}u_{m}\alpha_{m}^{+}\alpha_{-m}^{+})^{-1}=u_{m}%
^{-1}(1-\upsilon_{m}u_{m}\alpha_{m}^{+}\alpha_{-m}^{+})$\newline Replacing
this quantity in $I_{2}$\newline$I_{2}=\left\langle BCS\right\vert \alpha
_{-m}^{{}}\alpha_{m}^{{}}(\frac{\partial u_{m}}{\partial\beta_{i}}%
+\frac{\partial\upsilon_{m}}{\partial\beta_{i}}u_{m}^{2}\alpha_{m}^{+}%
\alpha_{-m}^{+})u_{m}^{-1}(1-\upsilon_{m}u_{m}\alpha_{m}^{+}\alpha_{-m}%
^{+})\left\vert BCS\right\rangle $\newline Again taking again into account
only two quasiparticle excitations,\newline$I_{2}=\left\langle BCS\right\vert
\alpha_{-m}^{{}}\alpha_{m}^{{}}(u_{m}\frac{\partial\upsilon_{m}}{\partial
\beta_{i}}-\upsilon_{m}\frac{\partial u_{m}}{\partial\beta_{i}})\alpha_{m}%
^{+}\alpha_{-m}^{+}\left\vert BCS\right\rangle $\newline we obtains:\newline%
$I_{2}=u_{m}\frac{\partial\upsilon_{m}}{\partial\beta_{i}}-\upsilon_{m}%
\frac{\partial u_{m}}{\partial\beta_{i}}$\newline knowing that the
normalization condition is:\newline$u_{m}^{2}+\upsilon_{m}^{2}=1$\newline we
find by differentiation\newline$2u_{m}\frac{\partial u_{m}}{\partial\beta_{i}%
}+2\upsilon_{m}\frac{\partial\upsilon_{m}}{\partial\beta_{i}}=0$\newline
combining these two relations, we obtain in $I_{2}$:\newline$I_{2}=-\frac
{1}{\upsilon_{m}}\frac{\partial u_{m}}{\partial\beta_{i}}$\newline then, the
second term reads:\newline$I_{2}=\left\langle m,-m\right\vert \left(
\frac{\partial}{\partial\beta_{i}}\right)  _{prob}\left\vert BCS\right\rangle
=-\frac{1}{\upsilon_{m}}\frac{\partial u_{m}}{\partial\beta_{i}}$\newline
which can be cast as follows:%

\begin{equation}
I_{2}=\left\langle k,-l\right\vert \left(  \frac{\partial}{\partial\beta_{i}%
}\right)  _{prob}\left\vert BCS\right\rangle =-\frac{1}{\upsilon_{k}}%
\frac{\partial u_{k}}{\partial\beta_{i}}\delta_{kl} \label{a4}%
\end{equation}
The two matrix elements $I_{1}$ and $I_{2}$ corresponding to the two cases
$k\neq l$ and $k=l$ are now known. Reassembling the two parts $I_{1}$ and
$I_{2}$ in the only one formula, we get:\newline$\left\langle M\right\vert
\frac{\partial}{\partial\beta_{i}}\left\vert O\right\rangle =I_{1}%
+I_{2}=\left\langle k,-l\right\vert \frac{\partial}{\partial\beta_{i}%
}\left\vert BCS\right\rangle =\left(  u_{k}\upsilon_{l}+u_{l}\upsilon
_{k}\right)  \left\langle k\right\vert \frac{\partial}{\partial\beta_{i}%
}\left\vert l\right\rangle \left(  1-\delta_{k,l}\right)  -\frac{1}%
{\upsilon_{k}}\frac{\partial u_{k}}{\partial\beta_{i}}\delta_{kl} $\newline
Replacing this quantity in the cranking formula, noting that the crossed terms
$(I_{1}I_{2}$ and $I_{2}I_{1})$ cancel in the product and knowing that
$E_{M}-E_{O}=E_{k}+E_{l}$ (see subsec.\ref{eground}) we find:%

\begin{equation}
D_{ij}\left\{  \beta_{1},.,\beta_{n}\right\}  =2\hbar^{2}\sum_{k,l}%
\frac{\left(  u_{k}\upsilon_{l}+u_{l}\upsilon_{k}\right)  ^{2}}{E_{k}+E_{l}%
}\left\langle l\right\vert \frac{\partial}{\partial\beta_{i}}\left\vert
k\right\rangle \left\langle k\right\vert \frac{\partial}{\partial\beta_{j}%
}\left\vert l\right\rangle \left(  1-\delta_{k,l}\right)  +2\hbar^{2}\sum
_{k}\frac{1}{2E_{k}}\frac{1}{\upsilon_{k}^{2}}\frac{\partial u_{k}}%
{\partial\beta_{i}}\frac{\partial u_{k}}{\partial\beta_{j}} \label{a5}%
\end{equation}
The first term of the r.h.s of the formula is the so-called non-diagonal term,
whereas the second is the diagonal one.

\section{Another version of the cranking formula\label{marker3}%
\label{appendix c}}

Let be $H$ some nuclear hamiltonian and $\left\vert O\right\rangle ,\left\vert
M\right\rangle $ its ground and excited state: noting that:\newline%
$\left\langle M\right\vert \left[  H,\left(  \frac{\partial}{\partial\beta
_{i}}\right)  _{wave\text{ }func}\right]  \left\vert O\right\rangle =\left(
E_{M}-E_{O}\right)  \left\langle M\right\vert \left(  \frac{\partial}%
{\partial\beta_{i}}\right)  _{wave\text{ }func}\left\vert O\right\rangle
$\newline since the commutator gives:\newline$\left[  H,\left(  \partial
/\partial\beta_{i}\right)  _{wave\text{ }func}\right]  =-\partial
H/\partial\beta_{i}$\newline the cranking formula becomes:\newline%
$D_{ij}\left\{  \beta_{1},.,\beta_{n}\right\}  =2\hbar^{2}\sum\limits_{M\neq
O}\frac{\left\langle O\right\vert \partial H/\partial\beta_{i}\left\vert
M\right\rangle \left\langle M\right\vert \partial H/\partial\beta
_{j}\left\vert O\right\rangle }{\left(  E_{M}-E_{O}\right)  ^{3}}+2\hbar
^{2}\sum\limits_{M\neq O}\tfrac{\left\langle O\right\vert \left(
\partial/\partial\beta_{i}\right)  _{occup.prob}\left\vert M\right\rangle
\left\langle M\right\vert \left(  \partial/\partial\beta_{j}\right)
_{occup.prob}\left\vert O\right\rangle }{\left(  E_{M}-E_{O}\right)  }%
$\newline where $\left(  \partial/\partial\beta_{i}\right)  _{wave\text{
}func}$ and $\left(  \partial/\partial\beta_{i}\right)  _{occup.prob}$ have
already defined in the appendix \ref{appendix b}.\newline In the
independent-particle approximation, we have in the second quantization
representation:\newline$\partial H/\partial\beta_{i}=\sum_{\nu,\mu
}\left\langle \nu\right\vert \partial H_{sp}/\partial\beta_{i}\left\vert
\mu\right\rangle a_{\nu}^{+}a_{\mu}^{{}}.$ After similar calculations than the
ones performed in the appendix \ref{appendix b}, we find\newline%
$I_{1}=\left\langle k,-l\right\vert \left(  \partial/\partial\beta_{i}\right)
_{wave\text{ }func}\left\vert BCS\right\rangle =-\left(  u_{k}\upsilon
_{l}+u_{l}\upsilon_{k}\right)  \left\langle k\right\vert \partial
H_{sp}/\partial\beta_{i}\left\vert l\right\rangle $ \ \ \ \ \ $k\neq
l$\newline therefore, the first part of the r.h.s of the above formula reads
(the second remains unchanged):

$2\hbar^{2}\sum\limits_{M\neq O}\frac{\left\langle O\right\vert \partial
H/\partial\beta_{i}\left\vert M\right\rangle \left\langle M\right\vert
\partial H/\partial\beta_{j}\left\vert O\right\rangle }{\left(  E_{M}%
-E_{O}\right)  ^{3}}=2\hbar^{2}\sum_{k,l}\frac{\left(  u_{k}\upsilon_{l}%
+u_{l}\upsilon_{k}\right)  ^{2}}{\left(  E_{k}+E_{l}\right)  ^{3}}\left\langle
l\right\vert \frac{\partial H_{sp}}{\partial\beta_{i}}\left\vert
k\right\rangle \left\langle k\right\vert \frac{\partial H_{sp}}{\partial
\beta_{j}}\left\vert l\right\rangle \left(  1-\delta_{k,l}\right)  $

Then we obtain for the cranking formula:%

\begin{equation}
D_{ij}\left\{  \beta_{1},.,\beta_{n}\right\}  =2\hbar^{2}\sum_{k,l}%
\frac{\left(  u_{k}\upsilon_{l}+u_{l}\upsilon_{k}\right)  ^{2}}{\left(
E_{k}+E_{l}\right)  ^{3}}\left\langle l\right\vert \frac{\partial H_{sp}%
}{\partial\beta_{i}}\left\vert k\right\rangle \left\langle k\right\vert
\frac{\partial H_{sp}}{\partial\beta_{j}}\left\vert l\right\rangle \left(
1-\delta_{k,l}\right)  +2\hbar^{2}\sum_{k}\frac{1}{2E_{k}}\frac{1}%
{\upsilon_{k}^{2}}\frac{\partial u_{k}}{\partial\beta_{i}}\frac{\partial
u_{k}}{\partial\beta_{j}} \label{a7}%
\end{equation}

\section{Final version of the cranking formula\label{marker4}%
\label{appendix d}}

The expression\newline$\sum_{k}\frac{1}{2E_{k}}\frac{1}{\upsilon_{k}^{2}}%
\frac{\partial u_{k}}{\partial\beta_{i}}\frac{\partial u_{k}}{\partial
\beta_{j}}$\newline meet in the above formula can be further clarified.
Recalling that:\newline$u_{k}^{{}}=\left(  1/\sqrt{2}\right)  \left(
1+\varepsilon_{k}/\sqrt{\varepsilon_{k}^{2}+\Delta^{2}}\right)  ^{1/2}$ and
$\upsilon_{k}^{{}}=\left(  1/\sqrt{2}\right)  \left(  1-\varepsilon_{k}%
/\sqrt{\varepsilon_{k}^{2}+\Delta^{2}}\right)  ^{1/2}$\newline
where:$\varepsilon_{k}=\epsilon_{k}-\lambda$ is the single-particle energy
with respect to the Fermi level and assuming for the moment that the
deformation dependence appears through $\epsilon_{k},\Delta,$ and $\lambda$,
we have to evaluate $\frac{\partial u_{k}}{\partial\beta}=\frac{\partial
u_{k}}{\partial\epsilon_{k}}\frac{\partial\epsilon_{k}}{\partial\beta}%
+\frac{\partial u_{k}}{\partial\lambda}\frac{\partial\lambda}{\partial\beta
}+\frac{\partial u_{k}}{\partial\Delta}\frac{\partial\Delta}{\partial\beta}$.
Thus, a simple differentiation of $u_{k}$ with respect to $\beta_{i}$ leads
to:\newline$\frac{\partial u_{k}^{{}}}{\partial\beta_{i}}=\frac{1}{2\sqrt{2}%
}\left(  1+\frac{\varepsilon_{k}}{\sqrt{\varepsilon_{k}^{2}+\Delta^{2}}%
}\right)  ^{-1/2}\left[  \frac{\partial\varepsilon_{k}}{\partial\beta_{i}%
}\left(  \varepsilon_{k}^{2}+\Delta^{2}\right)  ^{-1/2}-\varepsilon_{k}\left(
\varepsilon_{k}^{2}+\Delta^{2}\right)  ^{-3/2}\left(  \varepsilon_{k}%
\frac{\partial\varepsilon_{k}}{\partial\beta_{i}}+\Delta\frac{\partial\Delta
}{\partial\beta_{i}}\right)  \right]  $\newline multiplying by $\upsilon
_{k}^{-1}$and simplifying we get$:$\newline$\upsilon_{k}^{-1}\frac{\partial
u_{k}^{{}}}{\partial\beta_{i}}=\frac{1}{2\left(  \varepsilon_{k}^{2}%
+\Delta^{2}\right)  }\left\{  \Delta\frac{\partial\varepsilon_{k}}%
{\partial\beta_{i}}-\varepsilon_{k}\frac{\partial\Delta}{\partial\beta_{i}%
}\right\}  $\newline using $\varepsilon_{k}=\epsilon_{k}-\lambda$, we
obtain:\newline$\upsilon_{k}^{-1}\frac{\partial u_{k}^{{}}}{\partial\beta_{i}%
}=\frac{1}{2\left(  \varepsilon_{k}^{2}+\Delta^{2}\right)  }\left\{
\Delta\frac{\partial\epsilon_{k}}{\partial\beta_{i}}-\Delta\frac
{\partial\lambda}{\partial\beta_{i}}-\left(  \epsilon_{k}-\lambda\right)
\frac{\partial\Delta}{\partial\beta_{i}}\right\}  $\newline Moreover, noting
that:\newline$\frac{\partial\epsilon_{k}}{\partial\beta_{i}}=\left\langle
k\right\vert \frac{\partial H_{sp}}{\partial\beta_{i}}\left\vert
k\right\rangle $\newline we find:\newline$\upsilon_{k}^{-1}\frac{\partial
u_{k}^{{}}}{\partial\beta_{i}}=\frac{1}{2\left(  \varepsilon_{k}^{2}%
+\Delta^{2}\right)  }\left\{  \Delta\left\langle k\right\vert \frac{\partial
H_{sp}}{\partial\beta_{i}}\left\vert k\right\rangle -\Delta\frac
{\partial\lambda}{\partial\beta_{i}}-\left(  \epsilon_{k}-\lambda\right)
\frac{\partial\Delta}{\partial\beta_{i}}\right\}  $\newline the quasiparticle
energy is $E_{k}=\left(  \varepsilon_{k}^{2}+\Delta^{2}\right)  ^{1/2}$ so
that:\newline$\upsilon_{k}^{-1}\frac{\partial u_{k}^{{}}}{\partial\beta_{i}%
}=\frac{1}{2E_{k}^{2}}\left\{  \Delta\left\langle k\right\vert \frac{\partial
H_{sp}}{\partial\beta_{i}}\left\vert k\right\rangle -\Delta\frac
{\partial\lambda}{\partial\beta_{i}}-\left(  \epsilon_{k}-\lambda\right)
\frac{\partial\Delta}{\partial\beta_{i}}\right\}  $\newline putting:\newline%
$R_{i}^{k}=-$ $\left\langle k\right\vert \frac{\partial H_{sp}}{\partial
\beta_{i}}\left\vert k\right\rangle +\dfrac{\partial\lambda}{\partial\beta
_{i}}+\frac{\left(  \epsilon_{k}-\lambda\right)  }{\Delta}\dfrac
{\partial\Delta}{\partial\beta_{i}}$\newline the product of the similar terms
gives finally:\newline$I_{2}=\sum_{k}\frac{1}{2E_{k}}\frac{1}{\upsilon_{k}%
^{{}}}\frac{\partial u_{k}}{\partial\beta_{i}}\frac{1}{\upsilon_{k}^{{}}}%
\frac{\partial u_{k}}{\partial\beta_{j}}=\sum_{k}\frac{1}{2E_{k}}\Delta
\frac{R_{i}^{k}}{2E_{k}^{2}}\Delta\frac{R_{j}^{k}}{2E_{k}^{2}}$\newline The
cranking formula of the mass parameters becomes therefore:%

\begin{equation}
D_{ij}\left\{  \beta_{1},.,\beta_{n}\right\}  =2\hbar^{2}\sum_{k,l}%
\frac{\left(  u_{k}\upsilon_{l}+u_{l}\upsilon_{k}\right)  ^{2}}{\left(
E_{k}+E_{l}\right)  ^{3}}\left\langle l\right\vert \frac{\partial H_{sp}%
}{\partial\beta_{i}}\left\vert k\right\rangle \left\langle k\right\vert
\frac{\partial H_{sp}}{\partial\beta_{j}}\left\vert l\right\rangle \left(
1-\delta_{k,l}\right)  +2\hbar^{2}\sum_{k}\frac{\Delta^{2}}{8E_{k}^{5}}%
R_{i}^{k}R_{j}^{k} \label{a8}%
\end{equation}

\end{document}